\documentclass[twocolumn,english,aps,superscriptaddress,nofootinbib]{revtex4}
\usepackage[T1]{fontenc}
\usepackage[latin9]{inputenc}
\usepackage{color}
\usepackage{amsmath}
\usepackage{graphicx}
\usepackage{amssymb}

\makeatletter
\@ifundefined{textcolor}{}
{%
 \definecolor{BLACK}{gray}{0}
 \definecolor{WHITE}{gray}{1}
 \definecolor{RED}{rgb}{1,0,0}
 \definecolor{GREEN}{rgb}{0,1,0}
 \definecolor{BLUE}{rgb}{0,0,1}
 \definecolor{CYAN}{cmyk}{1,0,0,0}
 \definecolor{MAGENTA}{cmyk}{0,1,0,0}
 \definecolor{YELLOW}{cmyk}{0,0,1,0}
 }

\usepackage{graphicx}
\usepackage{bm}

\makeatother

\usepackage{babel}

\begin{document}

\title{Experimental criteria for steering and the Einstein-Podolsky-Rosen
paradox}

\author{E. G. Cavalcanti}

\affiliation{Centre for Quantum Dynamics, Griffith University, Brisbane QLD 4111,
Australia}

\affiliation{ARC Centre of Excellence for Quantum-Atom Optics, The University
of Queensland, Brisbane, QLD 4072, Australia}

\author{S. J. Jones}

\affiliation{Centre for Quantum Dynamics, Griffith University, Brisbane QLD 4111,
Australia}

\author{H. M. Wiseman}

\affiliation{Centre for Quantum Dynamics, Griffith University, Brisbane QLD 4111,
Australia}

\author{M. D. Reid}

\affiliation{ARC Centre of Excellence for Quantum-Atom Optics, The University
of Queensland, Brisbane, QLD 4072, Australia}

\date{\today{}}
\begin{abstract}
We formally link the concept of steering (a concept created by Schrödinger
but only recently formalised by Wiseman, Jones and Doherty {[}Phys.
Rev. Lett. 98, 140402 (2007){]} and the criteria for demonstrations
of Einstein-Podolsky-Rosen (EPR) paradox introduced by Reid {[}Phys.
Rev. A, 40, 913 (1989){]}. We develop a general theory of experimental
EPR-steering criteria, derive a number of criteria applicable to discrete
as well as continuous-variables observables, and study their efficacy
in detecting that form of nonlocality in some classes of quantum states.
We show that previous versions of EPR-type criteria can be rederived
within this formalism, thus unifying these efforts from a modern quantum-information
perspective and clarifying their conceptual and formal origin. The
theory follows in close analogy with criteria for other forms of quantum
nonlocality (Bell-nonlocality and entanglement), and because it is
a hybrid of those two, it may lead to insights into the relationship
between the different forms of nonlocality and the criteria that are
able to detect them.
\end{abstract}
\maketitle

\section{Introduction}

In their seminal 1935 paper \citep{Einstein1935}, Einstein, Podolsky
and Rosen (EPR) presented an argument which demonstrates the incompatibility
between the concepts of \emph{local causality} %
\footnote{This is Bell's terminology \citep{Bell1971}. It is also commonly
called local realism \citep{Reid1989}, which is arguably closer to
EPR's terminology. See however Ref.~\citep{Wiseman2006} for a discussion
of Einstein's later writings on locality and realism.%
} and the \emph{completeness} of quantum mechanics. Apart from the
foundational importance of that work, it had long-reaching consequences
\citep{Vedral2006}: it was the first time that physicists clearly
noticed the strange phenomena associated with \emph{entanglement}
--- the resource at the basis of modern quantum information science.

The situation depicted by EPR is often referred to as the ``EPR paradox''.
The authors themselves did not intend to point out a true paradox;
instead they argued that quantum mechanics was an incomplete theory,
that is, that it did not give a complete description of reality. Schrödinger
\citep{SchPCP35} seems to have been the first to name the situation
a `paradox', as he could not believe with EPR that quantum mechanics
was indeed incomplete but neither could he see a flaw in the argument.
In hindsight, we now know (since Bell \citep{Bell1964}) that, while
the argument is sound, one of the premises --- local causality ---
is false. However, we will retain the historically prevalent term
`paradox', if only because we still do not have a fully satisfactory
understanding of the nature of quantum nonlocality.

The original EPR paradox involved an example of an idealized bipartite
entangled state of continuous variables measured at the two subsystems.
Later, Bohm \citep{Bohm1951} extended the EPR paradox to a scenario
involving discrete (spin) observables. The essence of both of these
arguments involved perfect correlations, and therefore neither the
original EPR paradox nor Bohm's version could be directly tested in
the laboratory without additional assumptions. Criteria for the experimental
demonstration of the EPR paradox, which can be used in situations
with non-ideal states, have been derived for the continuous-variables
scenario by Reid in 1989 \citep{Reid1989} and more recently for discrete
systems by Cavalcanti and Reid \citep{Cavalcanti2007b} and Cavalcanti\emph{
et al. }\citep{Cavalcanti2009b}. 

In another recent development, Wiseman, Jones and Doherty \citep{Wiseman2007}
have introduced a new classification of quantum nonlocality, a formalisation
of the concept of \emph{steering} introduced by Schrödinger in 1935
\citep{Schroedinger1935} in a response to the EPR paper. In that
Letter, the authors claimed that any demonstration of the EPR paradox,
as proposed by Reid, is also a demonstration of steering. While that
claim was essentially correct, the proof proposed there was incomplete,
as we will see later in this paper. We will provide the missing proof
and further show that the converse is also true: any demonstration
of steering is also a demonstration of the EPR paradox. In other words,
the EPR paradox and steering are equivalent notions of nonlocality.

In Ref. \citep{Wiseman2007} Wiseman, Jones and Doherty showed that
EPR-steering constitutes a different class of nonlocality intermediate
between the classes of quantum non-separability and Bell-nonlocality,
with the distinction between these being explainable as a matter of
trust between different parties. Therefore, besides its foundational
interest, this classification could prove important in the context
of quantum communication and information. It would be thus desirable
to devise criteria to determine to which classes a given state (or
a set of observed correlations) belongs. For that purpose we will
formulate and develop the theory of \emph{EPR-steering criteria},
defined as any criteria which are sufficient to demonstrate EPR-steering
experimentally. The theory will proceed in close analogy to the theories
of entanglement criteria \citep{Duan2000,Simon2000,Hofmann2003a,Guhne2004a}
and of Bell inequalities\emph{ }(or Bell-nonlocality criteria)\emph{
}\citep{Bell1964,Clauser1969,Mermin1980,Fine1982,Pitowsky1989,Ardehali1992,Belinskii1993,Peres1999,Werner2001,Collins2002a,Zukowski2002a,Cavalcanti2007a}\emph{.}

The structure of the paper is as follows: In Sec.~\ref{sec:History-and-concepts}
we will review some of the history and concepts surrounding the EPR
paradox and steering. The main purposes of this section are to review
the conceptual motivation for the new formulation and to put the steering
criteria proposed here in context with the relevant literature. In
Sec.~\ref{sec:Locality-models} we will review the three classes
of nonlocality, including Wiseman and coworkers' \citep{Wiseman2007}
steering, and argue in more detail than in previous papers \citep{Jones2007}
as to why it provides the correct formalization of Schrödinger's concept.
In Sec.~\ref{sec:Experimental-criteria-for} we will introduce the
formalism for derivation of general EPR-steering criteria. We develop
two broad classes of EPR-steering criteria: the \emph{multiplicative
variance criteri}a, and the \emph{additive convex criteria} (which
includes linear EPR-steering inequalities as a special case). We show
how the criteria in the existing literature can be rederived as special
cases within this modern unifying approach. In Sec.~\ref{sec:Applications-to-classes}
we will apply the criteria derived in Sec.~\ref{sec:Experimental-criteria-for}
to some classes of quantum states, comparing their effectiveness in
experimentally demonstrating EPR-steering. We consider both continuous
variables (as in the original EPR paradox) and spin-half systems (as
in Bohm's version).

\section{History and concepts\label{sec:History-and-concepts}}

\subsection{The Einstein-Podolsky-Rosen argument\label{sec:The-Einstein-Podolsky-Rosen-argument}}

The EPR argument has been exhaustively commented in the literature.
However, since in this paper we will discuss a new mathematical formulation
of it, it will be important to review it in detail. 

The essence of Einstein and coworkers' \citep{Einstein1935} 1935
argument is a demonstration of the incompatibility between the premises
of \emph{local causality} and the \emph{completeness} of quantum mechanics.
EPR started the paper by making a distinction between \emph{reality}
and the \emph{concepts} of a theory, followed by a critique of the
operationalist position, clearly aimed at the views advocated by Bohr,
Heisenberg and the other proponents of the Copenhagen interpretation.
\begin{quote}
``Any serious consideration of a physical theory must take into account
the distinction between the objective reality, which is independent
of any theory, and the physical concepts with which the theory operates.
These concepts are intended to correspond with the objective reality,
and by means of these concepts we picture this reality to ourselves.

In attempting to judge the success of a physical theory, we may ask
ourselves two questions: (1) `Is the theory correct?' and (2) `Is
the description given by the theory complete?' It is only in the case
in which positive answers may be given to both of these questions,
that the concepts of the theory may be said to be satisfactory.''
\citep{Einstein1935}
\end{quote}
Any theory will have some concepts which will be used to aid in the
description and prediction of the phenomena which are their subject
matter. In quantum theory, Schrödinger introduced the concept of the
wave function and Heisenberg described the same phenomena with the
more abstract matrix mechanics. EPR argued that we must distinguish
those concepts from the reality they attempt to describe. One can
see the physical concepts of the theory as mere calculational tools
if one wishes, but it was those authors' opinion that one must be
careful to avoid falling back into a pure operationalist position;
the theory must strive to furnish a complete picture of reality.

EPR follow the previous considerations with a \emph{necessary condition
for completeness:}
\begin{quote}
\textbf{EPR's necessary condition for completeness: }``Whatever the
meaning assigned to the term \emph{complete, }the following requirement
for a complete theory seems to be a necessary one: \emph{every element
of the physical reality must have a counterpart in the physical theory}.''
\citep{Einstein1935}
\end{quote}
Soon afterward they note that this condition only makes sense if one
is able to decide what are the elements of the physical reality. They
did not attempt to \emph{define} `element of physical reality', saying
``The elements of the physical reality cannot be determined by \emph{a
priori }philosophical considerations, but must be found by an appeal
to results of experiments and measurements. A comprehensive definition
of reality is, however, unnecessary for our purpose''. Instead they
provide a \emph{sufficient condition}:
\begin{quote}
\textbf{EPR's sufficient condition for reality:} We shall be satisfied
with the following criterion, which we regard as reasonable. \emph{If,
without in any way disturbing a system, we can predict with certainty
(i.e., with probability equal to unity) the value of a physical quantity,
then there exists an element of physical reality corresponding to
this physical quantity}.'' \citep{Einstein1935}
\end{quote}
Later in the same paragraph it is made explicit that this criterion
is ``regarded not as a necessary, but merely as a sufficient, condition
of reality''. This is followed by a discussion to the effect that,
in quantum mechanics, if a system is in an eigenstate of an operator
$A$ with eigenvalue $a$, by this criterion, there must be an element
of physical reality corresponding to the physical quantity $A$. ``On
the other hand'', they continue, if the state of the system is a superposition
of eigenstates of $A$, ``we can no longer speak of the physical quantity
$A$ having a particular value''. After a few more considerations,
they state that ``the usual conclusion from this in quantum mechanics
is that \emph{when the momentum of a particle is known, its coordinate
has no physical reality}''\emph{. }We are left therefore, according
to EPR, with two alternatives:
\begin{quote}
\textbf{EPR's dilemma: }``From this follows that either (1) \emph{the
quantum-mechanical description of reality given by the wave function
is not complete or }(2) \emph{when the operators corresponding to
two physical quantities do not commute the two quantities cannot have
simultaneous reality.}'' \citep{Einstein1935}
\end{quote}
They justify this by reasoning that ``if both of them had simultaneous
reality --- and thus definite values --- these values would enter
into the complete description, according to the condition for completeness''.
And in the crucial step of the reasoning: ``If then the wave function
provided such a complete description of reality it would contain these
values; \emph{these would then be predictable \label{quo:EPR-predictable}}
{[}our emphasis{]}. This not being the case, we are left with the
alternatives stated''. Brassard and Méthot \citep{Brassard2006} (correctly)
pointed out that strictly speaking EPR should conclude that (1) \emph{or
}(2), instead of \emph{either} (1) or (2), since they could not exclude
the possibility that (1) and (2) could be both correct. However, this
does not affect EPR's conclusion. It was enough for them to show that
(1) and (2) could not both be wrong, and therefore if one can find
a reason for (2) to be false, (1) must be true %
\footnote{Brassard and Méthot's further conclusion that the EPR argument is
logically unsound is not based on this mistake, which they acknowledge
as irrelevant. Their conclusion is, in the present authors' opinion,
based on a misinterpretation of EPR's paper. They read the quote ``In
quantum mechanics it is usually assumed that the wave function \emph{does}
contain a complete description of the physical reality {[}...{]}.
We shall show however, that this assumption, together with the criterion
of reality given above, leads to a contradiction'', as stating that
$\neg(1)\wedge(2)\rightarrow false$. If that was the correct formalisation
of the argument we would agree with their conclusion. However, by
``criterion of reality given above'' EPR clearly mean their \textquotedbl{}sufficient
condition for reality\textquotedbl{}, not statement $(2)$.%
}.

The next section in EPR's paper intends to find a reason for (2) to
be false, that is, to find a circumstance in which one can say that
there are simultaneous elements of reality associated to two non-commuting
operators. They consider a composite system composed of two spatially
separated subsystems $S_{A}$ and $S_{B}$ which is prepared, by way
of a suitable initial interaction, in an entangled state of the type\begin{equation}
|\Psi\rangle=\sum_{n}c_{n}|\psi_{n}\rangle_{A}\otimes|u_{n}\rangle_{B},\label{eq:entangled1}\end{equation}
where the $|\psi_{n}\rangle_{A}$ denote a basis of eigenstates of
an operator, say $\hat{O}_{1}$, of subsystem $S_{A}$ and $|u_{n}\rangle_{B}$
denote some (normalised but not necessarily orthogonal) states of
$S_{B}$. If one measures the quantity $\hat{O}_{1}$ at $S_{A}$,
and obtains an outcome corresponding to eigenstate $|\psi_{k}\rangle_{A}$
the global state is reduced to $|\psi_{k}\rangle_{A}\otimes|u_{k}\rangle_{B}$.
If, on the other hand, one chooses to measure a non-commuting observable
$\hat{O}_{2}$, with eigenstates $|\phi_{s}\rangle_{A}$, one should
instead use the expansion\begin{equation}
|\Psi\rangle=\sum_{s}c'_{s}|\phi_{s}\rangle_{A}\otimes|v_{s}\rangle_{B},\label{eq:entangled2}\end{equation}
where $|v_{s}\rangle_{B}$ represent, in general, another set of states
of $S_{B}$. Now if the outcome of this measurement is, say, the one
corresponding to $|\phi_{r}\rangle_{A}$, the global state is thereby
reduced to $|\phi_{r}\rangle_{A}\otimes|v_{r}\rangle_{B}$. Therefore,
``as a consequence of two different measurements performed upon the
first system, the second system may be left in states with two different
wave functions''. This is just what Schrödinger later termed \emph{steering,
}and we will return to that later. Now enters the crucial assumption
of \emph{locality}, justified by the fact that the systems are spatially
separated and thus no longer interacting.
\begin{quote}
\textbf{EPR's necessary condition for locality: }``No real change
can take place in the second system in consequence of anything that
may be done to the first system.'' \citep{Einstein1935}
\end{quote}
Einstein \emph{et al.} never explicitly used the term `locality',
but took this assumption for granted. Because of this we call this
a ``necessary condition for locality'', as this is the most conservative
reading of EPR's reasoning: if they had explicitly defined some assumption
of locality, this would certainly be an implication of it, but there
is no reason (and no need) to take it as a definition. 

\begin{figure}
\begin{centering}
\includegraphics[width=8cm]{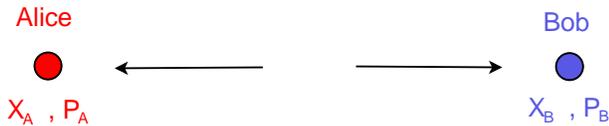}
\par\end{centering}

\caption{\label{fig:EPR} The EPR scenario. Alice and Bob are two spatially
separated observers who can perform one of two (position or momentum)
measurements available to each of them.}

\end{figure}

``\emph{Thus}'', conclude EPR, ``\emph{it is possible to assign two
different wave functions to the same} \emph{reality}''. EPR could
have now simply concluded by noting that two different (pure) states
can in general assign unit probability (and thus an element of reality,
according to the locality assumption and the sufficient condition
for reality) to each of two non-commuting quantities, in contradiction
of statement (2); this would imply, by way of EPR's dilemma, that
quantum mechanics is incomplete. Instead, they consider a specific
example, depicted in Fig.~\ref{fig:EPR}, where those different wave
functions are respective eigenstates of position and momentum. Because
they are canonically conjugate, this guarantees that $|u_{n}\rangle$
is different from $|v_{s}\rangle$ for \emph{every} possible outcome
\emph{n} or \emph{s}. The paradox is thus guaranteed to be realised
--- one cannot attempt to hide behind statistics. If the initial state
was of type\begin{equation}
\Psi(x_{A},x_{B})=\int_{-\infty}^{\infty}e^{ix_{A}p/\hbar}e^{-ix_{B}p/\hbar}dp,\label{eq:EPRstate}\end{equation}
then if one measures momentum $\hat{p}^{A}$ at $S_{A}$ and finds
outcome $p$, the reduced state of subsystem $S_{B}$ will be the
one associated with outcome $-p$ of $\hat{p}^{B}$. On the other
hand, if one measures position $\hat{x}^{A}$ and finds outcome $x$,
the reduced state of $S_{B}$ will be the one corresponding to outcome
$x$ of $\hat{x}^{B}$. By measuring position or momentum at $S_{A}$,
one can predict with certainty the outcome of the same measurement
on $S_{B}$. But $\hat{p}^{B}$ and $\hat{x}^{B}$ correspond to non-commuting
operators. EPR conclude from this that
\begin{quote}
``In accordance with our criterion of reality, in the first case we
must consider the quantity {[}$\hat{p}^{B}${]} as being an element
of reality, in the second case the quantity {[}$\hat{x}^{B}${]} is
an element of reality. But, as we have seen, both wave functions {[}corresponding
to $-p$ and $x${]} belong to the same reality.'' \citep{Einstein1935}
\end{quote}
In other words, by using the sufficient condition for reality, the
necessary condition for locality and the predictions for the entangled
state under consideration, EPR conclude that there must be elements
of reality associated to a pair of non-commuting operators. So horn
(2) of EPR's dilemma is closed, leaving as the only alternative option
(1), namely, that the quantum mechanical description of physical reality
is incomplete.

In more modern terminology, the conclusion of EPR was to infer the
existence of a set of local hidden variables (LHVs) underlying quantum
systems which should be able to reproduce the statistics. It is trivial
to reproduce the statistics of EPR's example with LHVs, even though
that is not possible with some entangled states, as later proved by
Bell \citep{Bell1964}. Schrödinger arrived at a different conclusion
from an analysis of the paradox raised by EPR, as we will see in the
next section.

In hindsight, as we now know that the premise of locality is not justified,
we can read EPR's argument as demonstrating the\emph{ incompatibility}
between the premises of locality, the completeness of quantum mechanics
and some of its predictions.

\subsection{Schrödinger's response: The concept of steering\label{sub:Schr=0000F6dinger's-response}}

EPR's argument prompted an interesting response from Schrödinger \citep{SchPCP35,Schroedinger1935}.
He also considered nonfactorizable pure states describable by the
wave function given by Eq. (\ref{eq:entangled1}). Schrödinger, however,
had of course developed the wave function for atoms and believed that
it gave a complete description of a quantum system. So while he was
not prepared to accept EPR's conclusion that quantum mechanics was
incomplete, neither could he see a flaw with their argument. For this
reason he termed the situation described by EPR a \emph{paradox}.

Clearly Schrödinger was also interested in implications arising from
composite quantum systems described by nonfactorizable pure states.
He described this situation, coining a famous term, as follows: {}``If
two separated bodies, each by itself known maximally, enter a situation
in which they influence each other, and separate again, then there
occurs regularly ... {[}an{]} entanglement of our knowledge of the
two bodies.'' \citep{SchPCP35}

Having defined entanglement, Schrödinger then defined the process
of \emph{disentanglement }which occurs when a non-degenerate observable
is measured on one body: {}``\emph{After establishing one representative
by observation, the other one can be inferred simultaneously ... this
procedure will be called} \emph{the} disentanglement''. This leads
us directly to the EPR paradox, as Schrödinger describes it:
\begin{quote}
{}``{[}EPR called attention{]} to the obvious but very disconcerting
fact that even though we restrict the disentangling measurements to
\emph{one} system, the representative obtained for the \emph{other}
system is by no means independent of the particular choice of observations
which we select for that purpose and which by the way are \emph{entirely}
arbitrary.''~\citep{SchPCP35}
\end{quote}
Schrödinger describes this ability to affect the state of the remote
subsystem as \emph{steering}:
\begin{quote}
{}``It is rather discomforting that the theory should allow a system
to be steered or piloted into one or the other type of state at the
experimenter's mercy in spite of his having no access to it.'' \citep{SchPCP35}
\end{quote}
EPR's example concerning position and momentum was recast in the context
of steering as
\begin{quote}
{}``Since I can predict \emph{either} $x_{1}$ or $p_{1}$ without
interfering with system No.~1 and since system No.~1, like a scholar
in examination, cannot possibly know which of the two questions I
am going to ask it first: it so seems that our scholar is prepared
to give the right answer to the \emph{first} question he is asked
\emph{anyhow}. He must know both answers; which is an amazing knowledge.''
\citep{SchPCP35}
\end{quote}
The remainder of Schrödinger's paper is a generalisation of steering
to more than two measurements:
\begin{quote}
``{[}System No.~1{]} does not only know these two answers but a vast
number of others, and that with no mnemotechnical help whatsoever,
at least none that we know of.'' \citep{SchPCP35}
\end{quote}
By {}``mnemotechnical help'' Schrödinger presumably means a cheat-sheet
(to use his scholar analogy). That is, a set of local hidden variables
(LHVs) that determine the measurement results. Thus, unlike EPR, Schrödinger
explicitly rejected LHVs as an explanation of steering. Perhaps because
he had performed explicit calculations generalizing EPR's example
(which can be explained trivially using LHVs), he recognized steering
as {}``a necessary and indispensable feature'' \citep{SchPCP36}
of quantum mechanics. We now know, thanks to Bell's theorem, that
Schrödinger's intuition was correct: there is no possible local hidden
variable model (or local mnemotechnical help) to explain the correlations
between measurement outcomes for certain entangled states \citep{Gisin1991}.

Like EPR, Schrödinger was troubled by the implications of steerability
of entangled states for quantum theory. Unlike EPR, however, he saw
the resolution of the paradox lying in the \emph{incorrectness} of
the predictions of quantum mechanics. That is, he was {}``\emph{not
satisfied about there being} \emph{sufficient experimental evidence
for}'' steering in nature \citep{SchPCP36}. This raises the obvious
question: what evidence would have convinced Schrödinger? The pure
entangled states he discussed are an idealization, so we cannot expect
ever to observe precisely the phenomenon he introduced. On the other
hand, Schrödinger was quite explicit that a separable but classically
correlated state which allows ``\emph{determining the state of the
first system by }suitable\emph{ measurement of the second or }vice
versa'' \citep{SchPCP36} could never exhibit steering. For this situation,
he says that ``\emph{it would utterly eliminate the experimenter's
influence on the state of that system which he does not touch}.''
\citep{SchPCP36}. Thus it is apparent that by steering Schrödinger
meant something that could not be explained by Alice simply finding
out which state Bob's system is in, out of some predefined ensemble
of states. Following this reasoning leads to the general definition
of steering as presented in Ref.~\citep{Wiseman2007}. We return
to this concept in Sec.~\ref{sec:Locality-models}.

\subsection{Bohm's version \label{sub:Bohm's-version}}

Although making reference to a general entangled state, the original
EPR argument used the specific case of a continuous-variable state
for its final (and crucial) part. In his 1951 textbook \citep{Bohm1951},
Bohm presented a discussion of the EPR paradox in a modified scenario
involving two entangled spin-1/2 particles. Although trivial in hindsight,
this extension had a fundamental importance. It was the scenario used
by Bell in the proof of his now famous theorem \citep{Bell1964} and
for most of the subsequent discussions of Bell inequalities (a Bell-type
inequality directly applicable to continuous-variables has only recently
been derived \citep{Cavalcanti2007a}), and was instrumental for our
present understanding of entanglement, and particularly for its applications
in quantum information processing. 

In Bohm's version the system of interest is a molecule containing
two spin-1/2 atoms in a singlet state, in which the total spin is
zero:\begin{equation}
|\Psi_{s}\rangle=|z_{+}\rangle_{A}\otimes|z_{-}\rangle_{B}-|z_{-}\rangle_{A}\otimes|z_{+}\rangle_{B}.\label{eq:Bohm_singlet_z}\end{equation}
Here $|z_{\pm}\rangle$ represent the $\pm1/2$ eigenstate of the
spin projection operator along the z direction, $S_{z}$. Compare
this state with Eq.~\eqref{eq:entangled1} used in the EPR argument.
If $S_{z}$ is measured on system \emph{A}, and the outcome corresponding
to $|z_{+}\rangle_{A}$ (or $|z_{-}\rangle_{A}$) is obtained, the
state of subsystem \emph{B} is projected into $|z_{-}\rangle_{B}$
(or $|z_{+}\rangle_{B}$). Thus, one predicts an element of reality
for the $z$ component of the spin of the second atom. But the same
state can be written, in the basis of eigenstates of another spin
projection, say $S_{x},$\begin{equation}
|\Psi_{s}\rangle=|x_{+}\rangle_{A}\otimes|x_{-}\rangle_{B}-|x_{-}\rangle_{A}\otimes|x_{+}\rangle_{B}.\label{eq:Bohm_singlet_x}\end{equation}
Similarly, the $x$ component of the spin of the first atom could
be measured instead, allowing inference of an element of reality associated
with the $x$ component of spin for the second atom. With this mapping,
the rest of the argument follows in analogy with EPR's. 

Bohm's version of the EPR paradox is conceptually appealing, but (in
his 1951 textbook at least) he did not present it as an argument for
the incompleteness of quantum theory (as did EPR). Instead, he used
it to argue that a complete description of nature need not contain
a one-to-one correspondence between elements of reality and the mathematical
description provided by the theory. Bohm defended, in 1951, the interpretation
that the quantum state represents only ``potentialities'' of measurement
results, which actually occur only when a system interacts with an
appropriate apparatus. It is curious to find that already in 1952
Bohm must have found this interpretation wanting, since he then developed
his famous non-local hidden-variable interpretation of quantum mechanics
\citep{Bohm1952a,Bohm1952b}, where there \emph{is} such a one-to-one
correspondence.

As the original continuous-variable example remained unrealizable
for decades, several early experiments followed Bohm's proposal, such
as Bleuler and Bradt (1948) \citep{Bleuler1948},\textbf{ }Wu and
Shaknov (1950) \citep{Wu1950} and Kocher and Commins (1967) \citep{Kocher1967}.
All of these suffered from low detection efficiencies and had no concern
with causal separation, however, making their interpretation debatable.

\subsection{The EPR-Reid criterion\label{sec:The-EPR-Reid-criterion}}

While the EPR argument was logically sound, one could block its conclusion
by rejecting those statistical predictions required to formulate it.
As we have discussed in Sec.~\ref{sub:Schr=0000F6dinger's-response},
Schrödinger seems to have found this an appealing solution. This move
is particularly easy to make since the necessary predictions are of
perfect correlations, unobtainable in practice due to unavoidable
inefficiency in preparation and detection of real physical systems.
This problem was considered by Furry already in 1936 \citep{Furry1936}
but experimentally useful criteria for the EPR paradox were only proposed
in 1989 by Reid \citep{Reid1989}, which we will discuss in detail
later in this section. The notation and terminology will closely follow
that of a recent review on the EPR paradox \citep{Reid2008tb}. The
essential difference in the derivation of the EPR-Reid criteria and
the original EPR argument is in a modification of the sufficient condition
for reality %
\footnote{Reid's original paper did not explicitly include this assumption,
which was implicit in the logic.%
}. This could be stated as the following:
\begin{quote}
\textbf{Reid's extension of EPR's sufficient condition of reality:}
If, without in any way disturbing a system, we can predict with \emph{some
specified uncertainty} the value of a physical quantity, then there
exists a \emph{stochastic} element of physical reality which determines
this physical quantity with at most that specific uncertainty.
\end{quote}
The scenario considered is the same as the one for the EPR paradox
above, as depicted in Fig.~\ref{fig:EPR}, but one does not need
a state which predicts the perfect correlations considered by EPR.
Instead, the two experimenters, Alice and Bob, can measure the conditional
probabilities of Bob finding outcome $x_{B}$ in a measurement of
$\hat{x}_{B}$ given that Alice finds outcome $x_{A}$ in a measurement
of $\hat{x}_{A}$, i.e., $P(x_{B}|x_{A})$. Similarly they can measure
the conditional probabilities $P(p_{B}|p_{A})$ and the unconditional
probabilities $P(x_{A})$, $P(p_{A})$. We denote by $\Delta^{2}(x_{B}|x_{A})$,
$\Delta^{2}(p_{B}|p_{A})$ the variances of the conditional distributions
$P(x_{B}|x_{A})$, $P(p_{B}|p_{A})$, respectively. Based on a result
$x_{A},$ Alice can make an estimate of the result for Bob's outcome
$x_{B}.$ Denote this estimate $x_{B}^{{\rm est}}(x_{A}).$ The\emph{
average inference variance} of $x_{B}$ given estimate $x_{B}^{{\rm est}}(x_{A})$
is defined as

\begin{multline}
\Delta_{\mathrm{inf}}^{2}x_{B}\equiv\langle(x_{B}-x_{B}^{{\rm est}}(x_{A}))^{2}\rangle\\
=\int dx_{A}dx_{B}P(x_{A},x_{B})(x-x_{B}^{{\rm est}}(x_{A}))^{2}.\label{eq:Inf_var_def}\end{multline}
Note that this average inference variance is minimized when the estimate
is just the expectation value of $x_{B}$ given $x_{A},$ i.e., the
mean of the distribution $P(x_{B}|x_{A})$ \citep{Reid2008tb}.\emph{
}Therefore the \emph{optimal (or minimum) inference variance} of $x_{B}$
($p_{B}$) given a measurement $\hat{x}_{A}$ ($\hat{p}_{A}$) is
given by \emph{\begin{eqnarray}
\Delta_{\mathrm{min}}^{2}x_{B} & = & \mathrm{min}_{x_{B}^{{\rm est}}}\{\Delta_{\mathrm{inf}}^{2}x_{B}\}\nonumber \\
 & = & \int dx_{A}dx_{B}P(x_{A})\Delta^{2}(x_{B}|x_{A});\label{eq:Reid_min_inf_var_x}\\
\Delta_{\mathrm{min}}^{2}p_{B} & = & \mathrm{min}_{p_{B}^{{\rm est}}}\{\Delta_{\mathrm{inf}}^{2}p_{B}\}\nonumber \\
 & = & \int dp_{A}dp_{B}P(p_{A})\Delta^{2}(p_{B}|p_{A}).\label{eq:Reid_min_inf_var_p}\end{eqnarray}
}Reid showed, by use of the sufficient condition of reality above,
that since Alice can, by measuring either position $\hat{x}_{A}$
or momentum $\hat{p}_{B}$, infer with some uncertainty $\Delta_{\mathrm{inf}}x_{B}=\sqrt{\Delta_{\mathrm{inf}}^{2}x_{B}}$
or $\Delta_{\mathrm{inf}}p_{B}=\sqrt{\Delta_{\mathrm{inf}}^{2}p_{B}}$
the outcomes of the corresponding experiments performed by Bob, and
since by the locality condition of EPR her choice cannot affect the
elements of reality of Bob, then there must be simultaneous stochastic
elements of reality which determine $\hat{x}_{B}$ and $\hat{p}_{B}$
with at most those uncertainties. Now by Heisenberg's Uncertainty
Principle (HUP), quantum mechanics imposes a limit to the precision
with which one can assign values to observables corresponding to non-commuting
operators such as $\hat{x}$ and $\hat{p}$. In appropriately rescaled
units the relevant HUP reads $\Delta x\Delta p\geq1$. Therefore,
if quantum mechanics is complete and the locality condition holds,
by use of the extended sufficient condition of reality and EPR's necessary
condition for completeness, the limit with which one could determine
the average inference variances above is\begin{equation}
\Delta_{\mathrm{inf}}x_{B}\Delta_{\mathrm{inf}}p_{B}\geq1.\label{eq:EPR-Reid}\end{equation}
This is the \emph{EPR-Reid criterion}. Violation of that criterion
signifies the EPR paradox, and has been experimentally demonstrated
in continuous-variables quantum optics experiments with quadratures
\citep{Ou1992,Zhang2000,Silberhorn2001,Schori2002,Bowen2003} and
actual position-momentum measurements \citep{Howell2004}. While these
were performed with high detection efficiency, none of these experimental
demonstrations have been able to achieve causal separation between
the measurements. For a detailed review see \citep{Reid2008tb}.

\subsection{Recent developments\label{sub:Recent-developments}}

Cavalcanti and Reid \citep{Cavalcanti2007b}\textbf{ }recently showed
that a larger class of quantum uncertainty relations can be used to
derive EPR inequalities. For example, from the uncertainty relation
$\Delta^{2}x+\Delta^{2}p\geq2,$ which follows from $\Delta x\Delta p\geq1,$
one can derive, in analogy with the previous section, the EPR criterion\begin{equation}
\Delta_{\mathrm{inf}}^{2}x_{B}+\Delta_{\mathrm{inf}}^{2}p_{B}\geq2.\label{eq:EPR_sum_Cav_Reid}\end{equation}
Using instead the spin uncertainty relation $\Delta J_{x}\Delta J_{y}\geq\frac{1}{2}|\langle J_{z}\rangle|,$
one can obtain the EPR criterion\begin{equation}
\Delta_{\mathrm{inf}}J_{x}^{B}\Delta_{\mathrm{inf}}J_{y}^{B}\geq\frac{1}{2}\sum_{J_{z}^{A}}P(J_{z}^{A})|\langle J_{z}^{B}\rangle_{J_{z}^{A}}|,\label{eq:EPR-Bohm_Cav_Reid}\end{equation}
useful for demonstration of Bohm's version of the EPR paradox. Here
$\langle J_{z}^{B}\rangle_{J_{z}^{A}}$ is the mean of the conditional
probability distribution $P(J_{z}^{B}|J_{z}^{A}).$ A weaker version
of Eq.~\eqref{eq:EPR-Bohm_Cav_Reid},\begin{equation}
\Delta_{\mathrm{inf}}J_{x}^{B}\Delta_{\mathrm{inf}}J_{y}^{B}\geq\frac{1}{2}|\langle J_{z}^{B}\rangle|,\label{eq:Bowen_EPR_criterion}\end{equation}
was used by Bowen \emph{et al. }\citep{Bowen2003} to demonstrate
an EPR paradox in the continuum limit for optical systems, with Stokes
operators playing the role of spin operators, in states where $\langle J_{z}^{B}\rangle\neq0.$

An inequality for demonstration of an EPR-Bohm paradox has also been
derived using an uncertainty relation based on sums of observables.
The uncertainty relation $\Delta^{2}J_{x}+\Delta^{2}J_{y}+\Delta^{2}J_{z}\geq\langle j\rangle,$
where $\langle j\rangle$ is the average total spin, has been used
in \citep{Hofmann2003a} for derivation of separability criteria,
and recently by \citep{Cavalcanti2009b} to derive the following EPR
criterion %
\footnote{More precisely, inequality \eqref{eq:sum_inf_var_criterion_J} was
presented in that work. The following follows with the substitution
explained below \eqref{eq:sum_inf_var_criterion_J}.%
}

\begin{equation}
\Delta_{\mathrm{inf}}^{2}J_{x}^{B}+\Delta_{\mathrm{inf}}^{2}J_{y}^{B}+\Delta_{\mathrm{inf}}^{2}J_{z}^{B}\geq\langle j^{B}\rangle.\label{eq:EPR-Bohm_sum_criterion}\end{equation}

All of the above EPR criteria will be rederived from an unifying perspective
in Section \ref{sec:Experimental-criteria-for}, and shown to be special
cases of broader classes of EPR-steering criteria.

\section{Locality models; EPR-steering\label{sec:Locality-models}}

In \citep{Wiseman2007}, a distinction was made between three locality
models, the failure of each corresponding to three strictly distinct
forms of nonlocality. To define those we will first establish some
notation.

Let $a\in\mathfrak{M_{\alpha}}$ and $b\in\mathfrak{M}_{\beta}$ represent
possible choices of measurements for two spatially separated observers
Alice and Bob, with respective outcomes denoted by the upper-case
variables $A\in\mathfrak{O}_{a}$ and $B\in\mathfrak{O}_{b}$, respectively.
Here we follow the case convention introduced by Bell \citep{Bell1964}.
Alice and Bob perform measurements on pairs of systems prepared by
a reproducible preparation procedure $c$. We denote the set of ordered
pairs $\mathfrak{M}\equiv\{(a,b):a\in\mathfrak{M}_{\alpha},b\in\mathfrak{M}_{\beta}\}$
a\emph{ measurement strategy}. The joint probability of obtaining
outcomes $A$ and $B$ upon measuring $a$ and $b$ after preparation
$c$ is denoted by\begin{equation}
P(A,B|a,b,c).\label{eq:jointProb}\end{equation}

The preparation procedure $c$ represents all those variables which
are explicitly known in the experimental situation. The joint probabilities
for all outcomes of all pairs of observables in a measurement strategy
given a preparation procedure define a \emph{phenomenon. }Following
Bell \citep{Bell1987}, we represent by $\lambda\in\Lambda$ any variables
associated with events in the union of the past light cones of $a,\, A,\, b,\, B$
which are\emph{ }relevant\emph{ }to the experimental situation but
are not explicitly known, and therefore not included in $c$. In this
sense they may be deemed hidden variables, but our usage will not
imply that they are necessarily hidden in principle (although in particular
theories they may be).

\subsection{Bell-nonlocality}

Given that notation, it is said that a phenomenon has a \emph{local
hidden variable} (\emph{LHV} or \emph{Bell-local }or \emph{locally
causal}) model if and only if for all $a\in\mathfrak{M_{\alpha}},\, A\in\mathfrak{O}_{a},\, b\in\mathfrak{M_{\beta}},\, B\in\mathfrak{O}_{b},$
there exist (i) a probability distribution $P(\lambda|c)$ over the
hidden variables, conditional on the information about the preparation
procedure $c$ %
\footnote{In general one could have a continuum of hidden variables, and Eq.
\eqref{eq:LHV} can be modified in the obvious way. No generality
is gained with that procedure, though, so we use the sum notation
for simplicity.%
} and (ii) arbitrary probability distributions $P(A|a,c,\lambda)$
and $P(B|b,c,\lambda)$, which reproduce the phenomenon in the form:
\begin{equation}
P(A,B|a,b,c)=\sum_{\lambda}P(\lambda|c)P(A|a,c,\lambda)P(B|b,c,\lambda).\label{eq:LHV}\end{equation}

Any constraint on the set of possible phenomena that can be derived
from \eqref{eq:LHV} is called\emph{ a Bell inequali}ty. A state for
which all phenomena can be given a LHV model, when the sets $\mathfrak{M_{\alpha}}$
and $\mathfrak{M_{\beta}}$ include all observables on the Hilbert
spaces of each corresponding subsystems, is called\emph{ a Bell-local
state}. If a state is not Bell-local it is called \emph{Bell-nonlocal}.

\subsection{Entanglement}

Similarly, it is said that a phenomenon has a \emph{quantum separable}
model, or \emph{separable} model for simplicity, if and only if for
all $a\in\mathfrak{M_{\alpha}},\, A\in\mathfrak{O}_{a},\, b\in\mathfrak{M_{\beta}},\, B\in\mathfrak{O}_{b},$
there exist $P(\lambda|c)$ as above and probability distributions
$P_{Q}(A|a,c,\lambda)$ and $P_{Q}(B|b,c,\lambda)$ such that

\begin{equation}
P(A,B|a,b,c)=\sum_{\lambda}P(\lambda|c)P_{Q}(A|a,c,\lambda)P_{Q}(B|b,c,\lambda),\label{eq:separable}\end{equation}
where now $P_{Q}(A|a,c,\lambda)$ represent probability distributions
for outcomes $A$ which are compatible with a quantum state. That
is, given a projector $\Pi_{a}^{A}$ associated to outcome $A$ of
measurement $a,$ and given a quantum density operator $\rho_{\alpha}(c,\lambda)$
for Alice's subsystem (as a function of $c$ and $\lambda$), these
probabilities are determined by \[
P_{Q}(A|a,c,\lambda)=\mathrm{Tr}\{\Pi_{a}^{A}\rho_{\alpha}(c,\lambda)\}.\]
Similar definitions apply for Bob's subsystem. 

Any constraint on the set of possible phenomena that can be derived
from assumption \eqref{eq:separable} is called a\emph{ separability
criterion} or\emph{ entanglement criterion}. A state for which all
phenomena can be given a separable model, when the sets $\mathfrak{M_{\alpha}}$
and $\mathfrak{M_{\beta}}$ include all observables on the Hilbert
spaces of each corresponding subsystems, is called a\emph{ separable
state}. A state which is not separable is called\emph{ non-separabl}e
o\emph{r entangled}. This definition is of course equivalent to the
usual definition involving product states, since if there is a separable
model for all possible measurement settings, then the joint state
can be given as a convex combination of product states \begin{equation}
\rho=\sum_{\lambda}P(\lambda|c)\rho_{\alpha}(c,\lambda)\otimes\rho_{\beta}(c,\lambda).\label{eq:separable_state}\end{equation}
Conversely, if the state is given as a convex combination of product
states of form \eqref{eq:separable_state}, the joint probabilities
for each pair of measurements are given straightforwardly by Eq.~\eqref{eq:separable}.

\subsection{EPR-steering}

Strictly intermediate between the LHV and separable models is the
\emph{local hidden-state (LHS) model for Bob.} This was argued in
\citep{Wiseman2007} to be the correct formalisation of non-steering
correlations. That is, violation of a LHS model for Bob is a demonstration
of EPR-steering, the concept introduced by Schrödinger to refer to
the situation depicted in the EPR paradox. Following the previous
notations, we say that a phenomenon has a \emph{no-Bob-steering model
}or a \emph{LHS model for Bob }(or\emph{ LHS model }for short) \emph{}%
\footnote{It would perhaps be more logical to use the term \emph{LHV/LHS model}
to denote no-steering, and the other types of nonlocality by LHV and
LHS models respectively, but we will use the simpler terminology introduced
in Ref.\citep{Wiseman2007}, as we believe there is no risk of confusion.%
} if and only if\emph{ }for all $a\in\mathfrak{M_{\alpha}},\, A\in\mathfrak{O}_{a},\, b\mathfrak{\in M_{\beta}},\, B\in\mathfrak{O}_{b},$
there exist $P(\lambda|c),$ $P(A|a,c,\lambda)$ and $P_{Q}(B|b,c,\lambda)$
defined as before such that

\begin{equation}
P(A,B|a,b,c)=\sum_{\lambda}P(\lambda|c)P(A|a,c,\lambda)P_{Q}(B|b,c,\lambda).\label{eq:LHS_model}\end{equation}

In other words, in a LHS model Bob's outcomes are described by some
quantum state, but Alice's outcomes are free to be arbitrarily determined
by the variables $\lambda.$ We call any constraint on the set of
possible phenomena that can be derived from \eqref{eq:LHS_model}
\emph{an EPR-steering criterion} \emph{or EPR-steering inequality}.
A state for which all phenomena can be given a LHS model, when the
sets $\mathfrak{M_{\alpha}}$ and $\mathfrak{M_{\beta}}$ include
all observables on the Hilbert spaces of each corresponding subsystems,
is called an\emph{ EPR-steerable state}. A state which is not steerable
is called\emph{ non-EPR-steerable}.

\subsection{Foundational relevance of EPR-steering}

As we have seen in Section \eqref{sub:Schr=0000F6dinger's-response},
Schrödinger was {}``discomforted'' with the possibility of Alice
being able to {}``steer'' Bob's system {}``in spite of {[}her{]}
having no access to it''. In other words, the strange phenomenon
revealed by the EPR paradox which he termed {}``steering'' was the
possibility that Alice could prepare, simply by different choices
of measurement on her own system, different ensembles of states for
Bob which are incompatible with a LHS model, that is, which cannot
be explained as arising from a coarse-graining from a pre-existing
ensemble of local quantum states for Bob. This is an inherently asymmetric
concept, thus the asymmetry in the formalization given by Eq.~\eqref{eq:LHS_model}.

For each choice of measurement $a,$ Alice will prepare for Bob one
state out of an ensemble $E^{a}\equiv\{\tilde{\rho}_{a}^{A}:A\in\mathfrak{O}_{a}\}.$
If the state of the global system is $\mathrm{W}_{c},$ the (unnormalized)
reduced state for Bob's subsystem corresponding to outcome $A$ will
be \begin{equation}
\tilde{\rho}_{a}^{A}\equiv\mathrm{Tr}_{\alpha}[\mathrm{W}_{c}(\Pi_{a}^{A}\otimes\mathbf{I})].\label{eq:Bob_red_state}\end{equation}
Evidently, the reduced density matrix for Bob is independent of Alice's
choice: $\rho_{\beta}=\mathrm{Tr}_{\alpha}[\mathrm{W}_{c}]=\sum_{A}\tilde{\rho}_{a}^{A}$
for all $a$ --- otherwise Alice could send faster-than-light signals
to Bob.

In Ref. \citep{Wiseman2007} it was shown that for pure states $\mathrm{W}_{c},$
entangled states, steerable states and Bell-nonlocal states are all
equivalent classes. The difficulty (and interest) comes when talking
about mixed states. In this case, one certainly does not want to consider
it as an example of steering when the ensembles prepared by Alice
are just different coarse-grainings of some underlying ensemble of
states. After all, these ensembles can be reproduced if Bob's local
state is simply classically correlated with some variables available
to Alice. These correlations would hardly constitute a puzzle for
Schrödinger, as we have argued in Section \eqref{sub:Schr=0000F6dinger's-response}.

Thus, Wiseman and co-workers \citep{Wiseman2007} considered EPR-steering
to occur iff it is not the case that there exists a decomposition
of Bob's reduced state, $\rho_{\beta}=\sum_{\lambda}P(\lambda|c)\rho_{\beta}(c,\lambda)$
such that for all $a\in\mathfrak{M_{\alpha}},\, A\in\mathfrak{O}_{a}$
there exists a stochastic map $P(A|a,c,\lambda)$ which allows all
states in the ensembles $E^{a}$ to be reproduced as \begin{equation}
\tilde{\rho}_{a}^{A}=\sum_{\lambda}P(A|a,c,\lambda)P(\lambda|c)\rho_{\beta}(c,\lambda).\label{eq:coarse-graining}\end{equation}

This definition leads directly to the formulation of a no-steering
model, Eq.~\eqref{eq:LHS_model}. According to the reduced state
\eqref{eq:coarse-graining}, the probability for outcome $B$ of Bob's
measurement $b,$ given an outcome $A$ of Alice's measurement $a,$
is given by $P(B|A,a,b,c)=\mathrm{Tr}[\Pi_{b}^{B}\tilde{\rho}_{a}^{A}]/P(A|a,b,c),$
where the denominator is introduced for normalization. Therefore the
joint probability becomes \begin{align}
P(A,B|a,b,c) & =\mathrm{Tr}[\Pi_{b}^{B}\tilde{\rho}_{a}^{A}]\nonumber \\
 & =\sum_{\lambda}P(A|a,c,\lambda)P(\lambda|c)\mathrm{Tr}[\Pi_{b}^{B}\rho_{\beta}(c,\lambda)]\nonumber \\
 & =\sum_{\lambda}P(\lambda|c)P(A|a,c,\lambda)P_{Q}(B|b,c,\lambda),\label{eq:LHS_model_again}\end{align}
as in Eq.~\eqref{eq:LHS_model}. The converse can also be trivially
shown.

One could propose that the definition of EPR-steering should take
into account the fact that Alice's state is also describable by quantum
mechanics. It can indeed be argued \citep{Cavalcanti2010tb} that
the conjunction of the assumptions of local causality and the completeness
of quantum mechanics (for both Alice and Bob) leads directly to a
quantum separable model, and in that sense EPR's conclusion that quantum
mechanics is incomplete (assuming local causality) could have been
reached by simply pointing out the predictions from any entangled
state. However, we are interested in capturing the phenomenon which
is central to EPR's \emph{actual} argument, and in Schrödinger's generalization
of this phenomenon, and hence we are led to the asymmetry in the definition.
This is the phenomenon that Einstein famously described as \textquotedbl{}spooky
action at a distance\textquotedbl{} \citep{Einstein1947}.

As we will see, this formalization also leads precisely to existing
EPR criteria, putting in a modern context the phenomena that have
already been discussed in the literature as generalizations of the
EPR paradox. Following Einstein's informal turn of phrase, we could
even call them tests of spooky action at a distance.

\subsection{EPR-steering as a quantum information task}

Wiseman and co-workers \citep{Wiseman2007,Jones2007} showed that
the distinction between the three forms of nonlocality above can be
formulated in a modern quantum information perspective, as a\emph{
task}. Suppose a third party, Charlie, wants proof that Alice and
Bob share an entangled state. Alice and Bob are not allowed to communicate,
but they can share any amount of classical randomness. If Charlie
trusts both Alice and Bob, he would be convinced iff Alice and Bob
are able to demonstrate entanglement, via violation of a separable
model, Eq.~\eqref{eq:separable}. If Charlie trusts Bob but not Alice,
he would be convinced they share entanglement iff they are able to
demonstrate EPR-steering by violating the local hidden state model
for Bob, Eq.~\eqref{eq:LHS_model}. If, on the other hand, Charlie
trusts neither of them, Alice and Bob would have to demonstrate Bell-nonlocality,
violating a local hidden variable model, Eq.~\eqref{eq:LHV}. The
reason is that, in the absence of trust, it is possible for the weaker
forms of nonlocality to be reproduced with the use of classical resources.

\section{Experimental criteria for EPR-steering\label{sec:Experimental-criteria-for}}

The above definition of EPR-steering invites the question: what are
the analogues for EPR-steering of Bell inequalities or entanglement
criteria, i.e., how can one derive what we have termed \emph{EPR-steering
criteria} above? In Refs. \citep{Wiseman2007} and \citep{Jones2007}
the emphasis was on the EPR-steering capabilities as a property of
states, and an analysis was made of how the steerability of some families
of quantum states depends on parameters which specify the states within
those families. This was necessary and useful for proving the strict
distinction between entangled, EPR-steerable and Bell-nonlocal states.
In an experimental situation, however, this kind of analysis is insufficient.
Quantum state tomography could be used to determine those parameters,
but what if the prepared state is only approximately a member of the
studied family? What about states which are not even approximately
members of any useful class? An experimental EPR-steering criterion
should not depend on any assumption about the type of state being
prepared, but only on the measured data. Compare this situation with
that of Bell inequalities, where a violation represents failure of
a LHV model, independently of any assumption about the state being
measured.

Another important issue is the relation between the EPR-type criteria
existing in the literature and the above formalization of EPR-steering.
In \citep{Wiseman2007} the authors provided a partial answer by showing
that for a class of Gaussian states the EPR-Reid criterion is violated
if and only if the state is steerable by Gaussian measurements. However,
the EPR-Reid criterion is valid for arbitrary states, and therefore
their conclusion that it is merely a special case of EPR-steering
was not entirely justified. Furthermore, the relation between this
formalization of EPR-steering and the other existing EPR-type criteria
cited in Sec.~\ref{sub:Recent-developments} was not discussed. Here
we will show that not only the EPR-Reid criterion but other existing
EPR-type criteria are indeed special cases of EPR-steering. We will
rederive those inequalities within this modern approach, and also
derive a number of new criteria for EPR-steering.

There is an important difference between Bell inequalities and EPR-steering
criteria. Since the LHV model \eqref{eq:LHV} does not depend on the
Hilbert space structure of quantum mechanics, Bell inequalities are
independent of the actual measurements being performed. To be clear,
the \emph{violation} of the inequality will certainly depend on which
measurements are performed (as well as the state being prepared),
but the derivation of the inequality itself is independent of that
information. In a Bell inequality the measurements are treated as
{}``black boxes'', where the only important feature is (usually,
but see \citep{Cavalcanti2007a}) their number of outcomes. In a LHS
model, on the other hand, Bob's subsytem is treated as a quantum state,
and therefore it is important in general to specify the actual quantum
operators corresponding to Bob's measurement choices, just as in an
entanglement criterion this information is in general required for
both Alice and Bob %
\footnote{The qualification 'in general' here is needed because a Bell inequality
\emph{is} an EPR-steering and an entanglement criterion. The failure
of a LHV model implies the failure of a LHS model and of a separable
model. However, in general a Bell inequality is inefficient as a criterion
for these weaker forms of nonlocality.%
}.

The fact that in a no-steering model Bob's probabilities are constrained
to be compatible with a quantum state suggests the use of quantum
uncertainty relations as ingredients in the derivation of criteria
for EPR-steering. A connection between uncertainty relations and EPR
criteria has been pointed out by two of the present authors in \citep{Cavalcanti2007b}
(although using the logic of the EPR-Reid criteria, not the present
formalization of EPR-steering), and that between uncertainty relations
and separability criteria has been shown by \citep{Hofmann2003a},
among others. 

We identify two main types of EPR-steering criteria: the \emph{multiplicative
variance} criteria, which include the EPR-Reid criteria and are based
on product uncertainty relations involving variances of observables;
and the \emph{additive convex} criteria, based on uncertainty relations
which are sums of convex functions.

\subsection{Existence of linear EPR-Steering criteria}

An interesting special case of additive convex criteria will be the
\emph{linear }criteria, based on linear functions of expectation values
of observables, and which can therefore be written as the expectation
value of a single Hermitian \emph{EPR-steering operator} $S$.

In general, for any (finite-dimensional) quantum state $W$, if the
state in question is steerable, then there exists a linear criterion
that would demonstrate EPR-steering for that phenomenon.

The proof is as follows. If the state is steerable, then by definition
there exists a measurement strategy which can demonstrate steering
with that state. Let $\mathfrak{M}$ be that measurement strategy.
Consider the set $\mathfrak{P(M)}$ of all possible phenomena for
$\mathfrak{M}$, i.e., the set of all possible sets of joint probabilities
$P(A,B|a,b)$ for all pair of outcomes $(A,B)$ of each pair of measurements
$(a,b)\in\mathfrak{M}$. Let $M$ be the number of possible settings
for the pair of measurements performed by Alice and Bob (i.e., the
number of elements in $\mathfrak{M}$) and let $O$ be the number
of possible pairs of outcomes $(A,B)$ for each pair of measurements.

A phenomenon is defined by specifying the $MO$ probabilities for
all possible outcomes of all measurements in the measurement strategy.
We represent those probabilities as an ordered set, and thus an element
$\mathbf{P}$ of $\mathfrak{P(M)}$ is associated to a point in $\mathbb{R}^{MO},$
where the joint probability for each $(A,B,a,b)$ is associated to
a coordinate $x_{ab}^{AB}$ of $\mathbb{R}^{MO}.$ For example, in
a phenomenon with 2 measurements per site with 2 outcomes each, $M=O=4,$
and the number of probabilities to be specified is $MO=16.$ Denoting
those measurements by $a\in\{a_{1},a_{2}\}$ and the outcomes of each
measurement by $A\in\{0,1\}$ (and similarly for Bob), these probabilities
would be represented by the vector $\mathbf{P}=(P(0,0|a_{1},b_{1}),P(0,1|a_{1},b_{1}),...,P(1,1|a_{2},b_{2})).$

Now consider two phenomena associated to $\mathbf{P}_{1}$ and $\mathbf{P}_{2}$,
and take a convex combination of the two vectors, i.e., \begin{equation}
\mathbf{P}_{3}=p\mathbf{P}_{1}+(1-p)\mathbf{P}_{2},\label{eq:convex_comb}\end{equation}
where $0\leq p\leq1$. If $\mathbf{P}_{1}$ and $\mathbf{P}_{2}$
have a no-steering model, then $\mathbf{P}_{3}$ also does. The proof
is simple: by assumption we can write the joint probabilities given
by $\mathbf{P}_{1}$ and $\mathbf{P}_{2}$ in form \eqref{eq:LHS_model}.
Simple manipulation shows that Eq.~\eqref{eq:convex_comb} can also
be written in form \eqref{eq:LHS_model}, with $P_{3}(\lambda)=pP_{1}(\lambda)+(1-p)P_{2}(\lambda)$.
In other words, the set of phenomena $\mathfrak{NS}\mathfrak{(M)}\subset\mathfrak{P(M)}$
which do not demonstrate EPR-steering is a convex set. (The same is
also true, of course, for the other forms of nonlocality.)

Now consider a phenomenon $\mathbf{P}_{s}\in\mathfrak{P(M)}$ which
\emph{does} demonstrate EPR-steering. By definition it is not in $\mathfrak{NS}\mathfrak{(M)}$.
Since, as shown above, that is a convex set, we can invoke a well
known result from convex analysis: there exists a plane in $\mathbb{R}^{MO}$
separating $\mathbf{P}_{s}$ from points in $\mathfrak{NS}\mathfrak{(M)}.$
Denote by $\hat{n}$ an unit vector normal to this plane pointing
away from $\mathfrak{NS}\mathfrak{(M)}$ and by $\mathbf{P}_{0}$
an arbitrary point on the plane. Then all points $\mathbf{P}_{\bar{s}}\in\mathfrak{NS}\mathfrak{(M)}$
satisfy \begin{equation}
\hat{n}\cdot(\mathbf{P}_{\bar{s}}-\mathbf{P}_{0})\leq0.\label{eq:steering_ineq_plane}\end{equation}
Inequality \eqref{eq:steering_ineq_plane} is an EPR-steering criterion.
If for an arbitrary point $\mathbf{P}_{c}\in\mathfrak{P(M)},$ $\hat{\mathbf{n}}\cdot(\mathbf{P}_{c}-\mathbf{P}_{0})>0,$
then $\mathbf{P}_{c}\notin\mathfrak{NS}$ and so this phenomenon demonstrates
EPR-steering. We can decompose $\mathbf{P}_{c}=\sum_{A,B,a,b}\langle\Pi_{a}^{A}\Pi_{b}^{B}\rangle_{c}\hat{\mathbf{e}}_{ab}^{AB},$
where $\langle\Pi_{a}^{A}\Pi_{b}^{B}\rangle_{c}\equiv P(A,B|a,b,c)=\mathrm{Tr}[W_{c}\,(\Pi_{a}^{A}\otimes\Pi_{b}^{B})]$
and $\{\hat{\mathbf{e}}_{ab}^{AB}\}$ is an orthonormal basis of $\mathbb{R}^{MO}$.
Decomposing $\hat{\mathbf{n}}=\sum_{A,B,a,b}n_{ab}^{AB}\hat{\mathbf{e}}_{ab}^{AB}$
and denoting $d\equiv-\hat{\mathbf{n}}\cdot\mathbf{P}_{0},$ \eqref{eq:steering_ineq_plane}
becomes $\sum_{A,B,a,b}n_{ab}^{AB}\langle\Pi_{a}^{A}\Pi_{b}^{B}\rangle_{c}+d\leq0.$
Defining a Hermitian operator $S\equiv\sum_{A,B,a,b}n_{ab}^{AB}\Pi_{a}^{A}\Pi_{b}^{B}+d\mathrm{I}$
we can rewrite the EPR-steering criterion \eqref{eq:steering_ineq_plane}
as \begin{equation}
\mathrm{Tr}[W_{c}S]\leq0,\label{eq:steer_operator_ineq}\end{equation}
which completes the proof.

However, this is merely an existence proof. It is quite a different
matter to produce the EPR-steering operator $S$ which will demonstrate
EPR-steering for a given state $W_{c}$. This is analogous to the
situation with Bell inequalities and entanglement, where one can prove
the existence of a Bell operator or entanglement witness for states
which can demonstrate the corresponding form of nonlocality, but cannot
easily produce such operators beyond some simple cases. 

Furthermore, in the case of EPR-steering (and also of entanglement)
the matter is even more complicated: there is an infinite (and continuous)
number of extreme points in the convex set of phenomena which allow
a LHS model (or a separable model) --- the set is not a polytope.
Therefore even for a finite measurement strategy, an infinite number
of linear inequalities are needed to fully specify the set. So in
general nonlinear criteria may be more useful, and we will consider
that general case in this paper. 

In the following subsections we will first derive the class of multiplicative
variance criteria, which will reduce to the well-known EPR-Reid criterion
as a special case. Then we will introduce the quite general class
of additive convex criteria, a special case of which will be the linear
criteria.

\subsection{Multiplicative variance criteria\label{sub:Multiplicative-variance-criteria}}

Following \citep{Reid1989}, we consider a situation where Alice tries
to infer the outcomes of Bob's measurements through measurements on
her subsystem. We denote by $B_{\mathrm{est}}(A)$ Alice's estimate
of the value of Bob's measurement $b$ as a function of the outcomes
of her measurement $a.$ As in Section \ref{sec:The-EPR-Reid-criterion},
the average inference variance of $B$ given estimate $B_{\mathrm{est}}(A)$
is defined by \begin{equation}
\Delta_{\mathrm{inf}}^{2}B=\langle(B-B_{\mathrm{est}}(A))^{2}\rangle.\label{eq:inf_var}\end{equation}
Here the average is over all outcomes $B$, $A$. Since for a given
$A$, the estimate that minimizes $\langle(B-B_{\mathrm{est}}(A))^{2}\rangle$
is just the mean $\langle B\rangle_{A}$ of \textcolor{black}{the
conditional probability $P(B|A),$} the optimal estimate for each
$A$ is just $B_{\mathrm{est}}(A)=\langle B\rangle_{A}.$ We denote
thus the \emph{optimal inference variance} of $B$ by measurement
of $a$ as \begin{eqnarray}
{\normalcolor \Delta_{\mathrm{min}}^{2}B} & = & \sum_{A,B}P(A,B)(B-\langle B\rangle_{A})^{2}\nonumber \\
 & = & \sum_{A}P(A)\sum_{B}P(B|A)(B-\langle B\rangle_{A})^{2}\nonumber \\
 & = & \sum_{A}P(A)\Delta^{2}(B|A)\label{eq:min_inf_var}\end{eqnarray}
where $\Delta^{2}(B|A)$ is the variance of $B$ calculated from the
conditional probability distribution $P(B|A).$ As explained above,
\begin{equation}
\Delta_{\mathrm{inf}}^{2}B\geq\Delta_{\mathrm{min}}^{2}B\label{eq:min<inf}\end{equation}
 for all choices of $B_{\mathrm{est}}(A).$ This minimum is optimal,
but not always experimentally accessible, in EPR experiments, since
it requires one to be able to measure conditional probability distributions. 

We assume that the statistics of Alice's and Bob's experimental outcomes
can be described by a LHS model, i.e., by a model of form \eqref{eq:LHS_model}
{[}omitting henceforth, for notational simplicity, the preparation
$c$ and the measurement choices $a,\, b$ from the conditional probabilities
$P(A,B|a,b,c),$ etc.{]}, \begin{equation}
P(A,B)=\sum_{\lambda}\, P(\lambda)\, P(A|\lambda)P_{Q}(B|\lambda).\label{eq:LHS model_2}\end{equation}
Assuming this model, the conditional probability of $B$ given $A$
is\begin{eqnarray}
P(B|A) & = & \sum_{\lambda}\frac{P(\lambda)P(A|\lambda)}{P(A)}P_{Q}(B|\lambda)\nonumber \\
 & = & \sum_{\lambda}P(\lambda|A)P_{Q}(B|\lambda).\label{eq:P(B|A)}\end{eqnarray}
As in Section \ref{sec:Locality-models}, $P_{Q}(B|\lambda)=\mathrm{Tr}[\Pi_{b}^{B}\rho_{\lambda}]$
represents the probability for $B$ predicted by a quantum state $\rho_{\lambda}.$
It is a general result that if a probability distribution has a convex
decomposition of the type $P(x)=\sum_{y}P(y)P(x|y),$ then the variance
$\Delta^{2}x$ over the distribution $P(x)$ cannot be smaller than
the average of the variances over the component distributions $P(x|y),$
\emph{i.e.}, $\Delta^{2}x\geq\sum_{y}P(y)\Delta^{2}(x|y).$ Therefore,
by \eqref{eq:P(B|A)}, the variance $\Delta^{2}(B|A)$ satisfies\begin{equation}
\Delta^{2}(B|A)\geq\sum_{\lambda}P(\lambda|A)\Delta_{Q}^{2}(B|\lambda),\label{eq:bound_var(B|A)}\end{equation}
where $\Delta_{Q}^{2}(B|\lambda)$ is the variance of $P_{Q}(B|\lambda).$
Using this result, we can derive a bound for Eq. \eqref{eq:min_inf_var},
\begin{equation}
\Delta_{\mathrm{min}}^{2}B\geq\sum_{A,\lambda}P(A,\lambda)\Delta_{Q}^{2}(B|\lambda)=\sum_{\lambda}P(\lambda)\Delta_{Q}^{2}(B|\lambda).\label{eq:bound_inf_var}\end{equation}

Suppose Bob's set of measurements consists of $\mathfrak{M}_{\beta}=\{b_{1},b_{2},b_{3}\},$
with respective outcomes labeled by $B_{1},\, B_{2},\, B_{3}.$ Alice
measures $\mathfrak{M}_{\alpha}=\{a_{1},a_{2},a_{3}\}.$ Suppose the
corresponding quantum observables for Bob, $\{\hat{b}_{1},\hat{b}_{2},\hat{b}_{3}\},$
obey the commutation relation $[\hat{b}_{1},\hat{b}_{2}]=i\hat{b}_{3}.$
The outcomes must then satisfy the product uncertainty relation \begin{equation}
\Delta_{Q}(B_{1}|\rho)\Delta_{Q}(B_{2}|\rho)\geq\frac{1}{2}|\langle B_{3}\rangle_{\rho}|,\label{eq:mult_UR}\end{equation}
where $\Delta_{Q}(B_{i}|\rho)$ and $\langle B_{i}\rangle_{\rho}$
are respectively the standard deviation and the average of $B_{i}$
in the quantum state $\rho.$ 

We will use the uncertainty relation above and the Cauchy-Schwarz
(C-S) inequality to obtain an EPR-steering criterion. The C-S inequality
states that, for two vectors $u$ and $v,$ $|u||v|\geq|u\cdot v|.$
Define $u=(\sqrt{P(\lambda_{1})}\Delta_{Q}(B_{1}|\lambda_{1})),\,\sqrt{P(\lambda_{2})}\Delta_{Q}(B_{1}|\lambda_{2}),\,\ldots)$
and $v=(\sqrt{P(\lambda_{1})}\Delta_{Q}(B_{2}|\lambda_{1}),\,\sqrt{P(\lambda_{2})}\Delta_{Q}(B_{2}|\lambda_{2}),\,\ldots).$
Then by \eqref{eq:bound_inf_var}\begin{eqnarray}
\Delta_{\mathrm{min}}B_{1}=\sqrt{\Delta_{\mathrm{min}}^{2}B_{1}} & \geq & |u|,\nonumber \\
\Delta_{\mathrm{min}}B_{2}=\sqrt{\Delta_{\mathrm{min}}^{2}B_{2}} & \geq & |v|.\label{eq:mult_var_u,v}\end{eqnarray}
We thus obtain, from \eqref{eq:mult_var_u,v}, the C-S inequality
and the uncertainty relation \eqref{eq:mult_UR}, \begin{eqnarray}
\Delta_{\mathrm{min}}B_{1}\Delta_{\mathrm{min}}B_{2} & \geq & |u||v|\nonumber \\
 & \geq & |u\cdot v|\nonumber \\
 & = & \sum_{\lambda}P(\lambda)\Delta_{Q}(B_{1}|\lambda)\Delta_{Q}(B_{2}|\lambda)\nonumber \\
 & \geq & \frac{1}{2}\sum_{\lambda}P(\lambda)|\langle B_{3}\rangle_{\lambda}|.\label{eq:mult_var_step2}\end{eqnarray}
Here we denote by $\langle B\rangle_{\lambda}$ the expectation value
of $B$ calculated from $P_{Q}(B|\lambda)$. Using again Eq.~\eqref{eq:P(B|A)}
and the fact that $f(x)=|x|$ is a convex function, that is, that
$\sum_{x}P(x)|x|\geq|\sum_{x}P(x)\, x|,$ we obtain a bound for the
last term: \begin{eqnarray}
\sum_{\lambda}P(\lambda)|\langle B_{3}\rangle_{\lambda}| & = & \sum_{A_{3},\lambda}P(A_{3},\lambda)|\langle B_{3}\rangle_{\lambda}|\nonumber \\
 & \geq & \sum_{A_{3}}P(A_{3})\left|\sum_{\lambda}P(\lambda|A_{3})\langle B_{3}\rangle_{\lambda}\right|\nonumber \\
 & = & \sum_{A_{3}}P(A_{3})|\langle B_{3}\rangle_{A_{3}}|\nonumber \\
 & \equiv & |\langle B_{i}\rangle|_{\mathrm{inf}}\label{eq:inf_ave}\end{eqnarray}
Using now \eqref{eq:min<inf}, we obtain, from \eqref{eq:mult_var_step2}
and \eqref{eq:inf_ave}, the EPR-steering criterion\begin{equation}
\Delta_{\mathrm{inf}}B_{1}\Delta_{\mathrm{inf}}B_{2}\geq\frac{1}{2}|\langle B_{3}\rangle|_{\mathrm{inf}}.\label{eq:mult_var_final}\end{equation}
This inequality was introduced in \citep{Cavalcanti2007b}, but its
derivation was based on the conceptual scheme of the EPR-Reid criterion.
Here we have shown that it follows directly from the LHS model \eqref{eq:LHS model_2}.
Its experimental violation implies the failure of the LHS model to
represent the measurement statistics, that is, it is an experimental
demonstration of EPR-steering. It is important to note that the choices
of measurement $a_{1},\, a_{2},\, a_{3}$ used by Alice to infer the
values of the corresponding measurements of Bob are arbitrary in this
derivation; the specific quantum observables $\hat{a}_{i}$ played
no role in the above because in a LHS model Alice's probabilities
are allowed to depend arbitrarily on the variables $\lambda$. In
an experimental situation, one should choose, of course, those which
can maximise the violation of \eqref{eq:mult_var_final}. 

One can also derive criteria involving collective variances such as
$\Delta^{2}(g_{k}A_{k}+B_{k}),$ where $g_{k}$ is a real number.
These measurements are often simpler to be realised as they do not
require the full conditional distributions. These are just the average
inference variances $\Delta_{\mathrm{inf}}^{2}B_{k}=\langle[B_{k}-B_{\mathrm{est}}(A_{k})]^{2}\rangle$
with a linear estimate $B_{\mathrm{est}}(A_{k})=-g_{k}A_{k}+\langle B_{k}+g_{k}A_{k}\rangle,$
as shown in \citep{Reid2008tb}. We can therefore straightforwardly
derive, from \eqref{eq:mult_var_final}:

\begin{equation}
\Delta(g_{1}A_{1}+B_{1})\Delta(g_{2}A_{2}+B_{2})\geq\frac{1}{2}|\langle B_{3}\rangle|_{\mathrm{inf}},\label{eq:mult_collective_var}\end{equation}
keeping in mind that the measurements for Alice and the values of
$g_{k}$ are arbitrary, and should be chosen so as to optimize the
violation of the inequality.

\subsubsection{Examples}

The first example of a multiplicative variance criterion is the original
EPR-Reid criterion \citep{Reid1989}, reviewed in Section \ref{sec:The-EPR-Reid-criterion}.
It was developed for continuous variables observables $\hat{x}^{B}$
and $\hat{p}^{B},$ which obey an uncertainty relation $\Delta_{Q}(x^{B}|\rho)\Delta_{Q}(p^{B}|\rho)\geq1,$
arising from the commutation relation (in appropriate units) $[\hat{x}^{B},\hat{p}^{B}]=2i$.
Substituting $B_{1}=x^{B},$ $B_{2}=p^{B}$ and $B_{3}=2$ in \eqref{eq:mult_var_final}
we obtain the EPR-Reid criterion \eqref{eq:EPR-Reid},

\begin{equation}
\Delta_{\mathrm{inf}}x^{B}\Delta_{\mathrm{inf}}p^{B}\geq1.\label{eq:EPR-Reid_rederived}\end{equation}

This provides a formal proof of the incomplete conjecture put forth
in \citep{Wiseman2007}, that the EPR-Reid criterion is a special
case of EPR-steering. It is a direct consequence of the assumption
of a LHS model; in particular this derivation does not require Reid's
extension of EPR's necessary condition for reality.

For angular momentum observables, obeying a commutation relation $[\hat{J}_{x}^{B},\hat{J}_{y}^{B}]=i\hat{J}_{z}^{B}$
(and its cyclical permutations) the corresponding quantum uncertainty
relation is $\Delta_{Q}(J_{x}^{B}|\rho)\Delta_{Q}(J_{y}^{B}|\rho)\geq\frac{1}{2}|\langle J_{z}^{B}\rangle_{\rho}|$
(and permutations). Substituting these in \eqref{eq:mult_var_final},
with $B_{1}=J_{x}^{B},$ $B_{2}=J_{y}^{B}$ and $B_{3}=J_{z}^{B},$
we obtain the criterion \eqref{eq:EPR-Bohm_Cav_Reid} reviewed in
Section \ref{sub:Recent-developments}:\begin{equation}
\Delta_{\mathrm{inf}}J_{x}^{B}\Delta_{\mathrm{inf}}J_{y}^{B}\geq\frac{1}{2}|\langle J_{z}^{B}\rangle|_{\mathrm{inf}},\label{eq:EPR-Bohm_criterion}\end{equation}
and of course, its permutations. Violation of one of these inequalities
corresponds to a demonstration of the EPR-Bohm paradox discussed in
Sec. \ref{sub:Bohm's-version}. Bowen \emph{et al.}'s \citep{Bowen2003}
inequality \eqref{eq:Bowen_EPR_criterion} is the special case in
which Alice's choice of measurement used to infer $|\langle J_{z}^{B}\rangle|_{\mathrm{inf}}$
is the identity. We can see that it is a weaker criterion than the
above by noting that the convexity of the function $f(x)=|x|$ implies
$|\langle J_{z}^{B}\rangle|_{\mathrm{inf}}\equiv\sum_{J_{z}^{A}}P(J_{z}^{A})|\langle J_{z}^{B}\rangle_{J_{z}^{A}}|\geq|\langle J_{z}^{B}\rangle|.$
Inequality \eqref{eq:Bowen_EPR_criterion} therefore will be violated
only if \eqref{eq:EPR-Bohm_criterion} also is. In particular, \eqref{eq:EPR-Bohm_criterion}
can detect EPR-steering for states in which the expectation value
of $J_{z}^{B}$ is zero, such as the symmetric state originally considered
by Bohm \citep{Bohm1951}. Applications of these criteria to specific
classes of quantum states will be given in Sec. \ref{sec:Applications-to-classes}.

\subsection{Additive convex criteria}

We now present the derivation of the class of additive convex criteria.
Suppose one has an uncertainty relation in the broadest sense ---
a general constraint which must be obeyed by all quantum states of
Bob's subsystem --- of form

\begin{equation}
\sum_{j}f_{j}(\langle B_{j}\rangle_{\rho},\alpha_{j})\leq0,\label{eq:add_conv_constraint}\end{equation}
where $j$ indexes observables on Bob's subsystem, $\langle B_{j}\rangle_{\rho}$
denotes the expectation value of observable $b_{j}$ on a quantum
state $\rho,$ $\alpha_{j}\in\mathbb{R}$ are parameters of the constraint
which can take any values in some set $\mathfrak{O}_{a_{j}}$ (the
significance of which should be clear soon), and the functions $f_{j}$
are convex on the interval containing the possible values of the first
argument (i.e., the possible expectation values $\langle B_{j}\rangle_{\rho}$,
which is the convex hull $H_{\mathrm{convex}}\{\mathfrak{O}_{b_{j}}\}$
of the set of possible outcomes of $b_{j}$). This last requirement
means that for all $x,\, y\in H_{\mathrm{convex}}\{\mathfrak{O}_{b_{j}}\},$
for all $z\in\mathfrak{O}_{a_{j}}$ and for all $p\in[0,1]$, \begin{equation}
f_{j}(px+(1-p)y,z)\leq pf_{j}(x,z)+(1-p)f_{j}(y,z).\end{equation}

Although the product uncertainty relations considered in the previous
section are not of form \eqref{eq:add_conv_constraint}, since they
include terms like $\langle B_{1}^{2}\rangle\langle B_{2}^{2}\rangle,$
a large class of uncertainty relations can be written in this form.
The negative of the variance of a variable $B$, that is, $-\Delta^{2}B=\langle B\rangle^{2}-\langle B^{2}\rangle,$
is a sum of two convex functions $f_{1}(\langle B\rangle)+f_{2}(\langle B^{2}\rangle),$
{[}with $f_{1}(x)=x^{2}$ and $f_{2}(x)=-x${]} and thus we can obtain
EPR-steering criteria from uncertainty relations that involve sums
of variances of observables. For example, the relation $\Delta^{2}B_{1}+\Delta^{2}B_{2}\geq|\langle B_{3}\rangle|$
\citep{Cohen-Tannoudji1977} can be rewritten as \begin{equation}
|\langle B_{3}\rangle|-\langle B_{1}^{2}\rangle+\langle B_{3}\rangle^{2}-\langle B_{3}^{2}\rangle+\langle B_{3}\rangle^{2}\leq0,\label{eq:UR_add_convex}\end{equation}
which is of form \eqref{eq:add_conv_constraint}, with 5 terms in
the sum. All terms are convex, since the coefficients of the square
terms and absolute-value terms are positive. Any term linear on the
expectation values $\langle B_{j}\rangle_{\rho}$ is clearly also
of that form. As in the previous section, the assumption that the
statistics of Alice and Bob can be described by a LHS model of form
\eqref{eq:LHS model_2} implies that the conditional probability of
outcome $B$ given outcome $A$ can be written as\begin{equation}
P(B|A)=\sum_{\lambda}P(\lambda|A)P_{Q}(B|\lambda).\label{eq:P(B|A)_again}\end{equation}
 The average of this conditional probability,$\langle B\rangle_{A},$
can be thus written as\begin{equation}
\langle B\rangle_{A}=\sum_{\lambda}P(\lambda|A)\langle B\rangle_{\lambda},\label{eq:ave_B_A}\end{equation}
and we remind the reader that $\langle B\rangle_{\lambda}\equiv\sum_{B}P_{Q}(B|\lambda)\, B=\mathrm{Tr}\{\hat{b}\,\rho_{\lambda}\}.$

If $f$ is a convex function, \eqref{eq:ave_B_A} then implies, for
all $A$,\begin{eqnarray}
f\left(\langle B\rangle_{A},A\right) & = & f\left(\sum_{\lambda}P(\lambda|A)\langle B\rangle_{\lambda},A\right)\nonumber \\
 & \leq & \sum_{\lambda}P(\lambda|A)\, f\left(\langle B\rangle_{\lambda},A\right).\label{eq:conv_f_step_1}\end{eqnarray}
Taking the average over $A$ we obtain\begin{equation}
\sum_{A}P(A)\, f\left(\langle B\rangle_{A},A\right)\leq\sum_{A,\lambda}P(A,\lambda)\, f\left(\langle B\rangle_{\lambda},A\right).\label{eq:ave_over_A}\end{equation}
 We now introduce the subscripts $j,$ sum both sides of \eqref{eq:ave_over_A}
over $j$ and apply the quantum constraint \eqref{eq:add_conv_constraint}
to obtain \begin{multline}
\sum_{j,A_{j}}P(A_{j})\, f_{j}\left(\langle B_{j}\rangle_{A_{j}},A_{j}\right)\\
\leq\sum_{A_{j},\lambda}P(A_{j},\lambda)\,\sum_{j}f_{j}\left(\langle B_{j}\rangle_{\lambda},A_{j}\right)\leq0\;.\end{multline}

Introducing the simplifying notation $E_{b|a}[f_{j}]\equiv\sum_{A_{j}}P(A_{j})\, f_{j}\left(\langle B_{j}\rangle_{A_{j}},A_{j}\right),$
we write the general EPR-steering criterion\begin{equation}
\sum_{j}E_{b|a}[f_{j}]\leq0\;.\label{eq:add_convex_criteria}\end{equation}
A weaker version of the inequality (i.e., one that detects steerability
less efficiently) can be obtained by using the following bound, which
is a consequence of the convexity of $f_{j},$ when $f_{j}$ does
not depend explicitly on $A_{j}$:\begin{equation}
f_{j}(\langle B_{j}\rangle)\leq E_{b|a}[f_{j}].\label{eq:meas_bound}\end{equation}
One can therefore substitute $E_{b|a}[f_{j}]$ by $f_{j}(\langle B_{j}\rangle)$
for some $j$ in \eqref{eq:add_convex_criteria} and the inequality
still holds.

\subsubsection{Examples: criteria from inference variances}

We will now give some examples of criteria that can be obtained with
the general form of \eqref{eq:add_convex_criteria}.We note, to make
contact with the previous notation, that when the $f_{j}$'s involve
variances, the corresponding expressions on the left-hand side of
\eqref{eq:add_convex_criteria} are just \begin{equation}
\sum_{A}P(A)\,\left(\langle B\rangle_{A}^{2}-\langle B^{2}\rangle_{A}\right)=-\Delta_{\mathrm{min}}^{2}B,\label{eq:fj_inf_var}\end{equation}
as defined on \eqref{eq:inf_var}. As before, the bound \begin{equation}
\Delta_{\mathrm{inf}}^{2}B\geq\Delta_{\mathrm{min}}^{2}B\label{eq:inf>min_again}\end{equation}
 can be used in the derivation of the inequalities.

We start considering arbitrary observables obeying commutation relation
$[\hat{b}_{1},\hat{b}_{2}]=i\hat{b}_{3},$ and use the uncertainty
relation $\Delta^{2}(B_{1}|\rho)+\Delta^{2}(B_{2}|\rho)\geq|\langle B_{3}\rangle_{\rho}|,$
which is of form \eqref{eq:add_conv_constraint} as shown above. Expanding
this in terms of the $f_{j}$'s, substituting on \eqref{eq:add_convex_criteria}
and using \eqref{eq:fj_inf_var} and \eqref{eq:inf>min_again} we
obtain the EPR-steering inequality\begin{equation}
\Delta_{\mathrm{inf}}^{2}B_{1}+\Delta_{\mathrm{inf}}^{2}B_{2}\geq|\langle B_{3}\rangle|_{\mathrm{inf}},\label{eq:sum_inf_var_criterion_arb}\end{equation}
where as before $|\langle B_{3}\rangle|_{\mathrm{inf}}\equiv\sum_{A_{3}}P(A_{3})|\langle B_{3}\rangle_{A_{3}}|,$
and the bound $|\langle B_{3}\rangle|_{\mathrm{inf}}\geq|\langle B_{3}\rangle|$
can be used if needed.

For continuous variables observables $[\hat{x}^{B},\hat{p}^{B}]=2i,$
\eqref{eq:sum_inf_var_criterion_arb} becomes inequality \eqref{eq:EPR_sum_Cav_Reid},
\begin{equation}
\Delta_{\mathrm{inf}}^{2}x^{B}+\Delta_{\mathrm{inf}}^{2}p^{B}\geq2,\label{eq:sum_inf_var_criterion_x}\end{equation}
and for angular momentum observables inequality \eqref{eq:sum_inf_var_criterion_arb}
reads \begin{equation}
\Delta_{\mathrm{inf}}^{2}J_{x}^{B}+\Delta_{\mathrm{inf}}^{2}J_{y}^{B}\geq|\langle J_{z}^{B}\rangle|_{\mathrm{inf}}.\label{eq:sum_inf_var_criterion_|J|}\end{equation}
Inequality \eqref{eq:sum_inf_var_criterion_x} has been derived (within
the EPR-Reid formalism) in \citep{Cavalcanti2007b}. However, these
inequalities are weaker than the corresponding multiplicative variance
criteria: since for any pair of real numbers $x^{2}+y^{2}\geq2xy,$
inequality \eqref{eq:mult_var_final} directly implies \eqref{eq:sum_inf_var_criterion_arb}
and thus the latter can be violated only if the former is.

Another special case of additive convex criterion has been recently
derived in \citep{Cavalcanti2009b}.\textcolor{black}{{} Consider Schwinger
spin operators defined as \begin{eqnarray}
\hat{J}_{x}^{B} & = & \frac{1}{2}\left(\hat{b}_{-}\hat{b}_{+}^{\dagger}+\hat{b}_{-}^{\dagger}\hat{b}_{+}\right),\nonumber \\
\hat{J}_{y}^{B} & = & \frac{1}{2i}\left(\hat{b}_{-}\hat{b}_{+}^{\dagger}-\hat{b}_{-}^{\dagger}\hat{b}_{+}\right),\nonumber \\
\hat{J}_{z}^{B} & = & \frac{1}{2}\left(\hat{b}_{+}^{\dagger}\hat{b}_{+}-\hat{b}_{-}^{\dagger}\hat{b}_{-}\right),\nonumber \\
\hat{N}^{B} & = & \left(\hat{b}_{+}^{\dagger}\hat{b}_{+}+\hat{b}_{-}^{\dagger}\hat{b}_{-}\right),\label{eq:Schwinger}\end{eqnarray}
where $\hat{b}_{\pm}$ ar}e boson operators for two field modes of
Bob's subsystem, obeying commutation relations $[\hat{b}_{\pm},\hat{b}_{\pm}^{\dagger}]=1.$
Similar operators are defined for Alice\textcolor{black}{. The situation
of the EPR-Bohm setup is therefore extended with number measurements.}
We now use the quantum uncertainty relation \citep{Hofmann2003a}
\begin{equation}
\Delta^{2}(J_{x}^{B}|\rho)+\Delta^{2}(J_{y}^{B}|\rho)+\Delta^{2}(J_{z}^{B}|\rho)\geq\frac{1}{4}\Delta^{2}(N^{B}|\rho)+\frac{1}{2}\langle N^{B}\rangle_{\rho},\label{eq:HUPHofmann}\end{equation}
and rewrite it in the form of \eqref{eq:add_conv_constraint}, $-\Delta^{2}(J_{x}^{B}|\rho)-\Delta^{2}(J_{y}^{B}|\rho)-\Delta^{2}(J_{z}^{B}|\rho)+\langle N^{B}\rangle_{\rho}/2\leq0,$
dropping the positive but non-convex term $\Delta^{2}N^{B}/4.$ Substituting
this in \eqref{eq:add_convex_criteria}, and using \eqref{eq:fj_inf_var}
and \eqref{eq:inf>min_again}, we obtain: \begin{equation}
\Delta_{\mathrm{inf}}^{2}J_{x}^{B}+\Delta_{\mathrm{inf}}^{2}J_{y}^{B}+\Delta_{\mathrm{inf}}^{2}J_{z}^{B}\geq\frac{\langle N^{B}\rangle}{2}.\label{eq:sum_inf_var_criterion_J}\end{equation}
In the angular momentum basis $\{|j,m\rangle\},$ where $j(j+1)$
are the eigenvalues of $\hat{J}{}^{2}=(\hat{J}_{x}^{2}+\hat{J}_{y}^{2}+\hat{J}_{z}^{2})$
and $m$ are the eigenvalues of $\hat{J}_{z},$ the operator $\hat{N}/2$
corresponds to the {}``total angular momentum'' operator $\hat{J}_{T}=\sum_{j}j\sum_{m}|j,m\rangle\langle j,m|,$
i.e., the operator which has a spectral decomposition in terms of
projectors onto each subspace of constant $j,$ with corresponding
eigenvalues $j.$ %
\footnote{Note that the angular momentum-square operator $J{}^{2}$ is not the
square of this operator. Although they have the same eigenvectors,
the eigenvalues of $J{}^{2}$ are $j(j+1)$ and not $j^{2}.$%
} Any criteria in which $\langle N^{B}\rangle$ occurs can therefore
be modified by substituting $\langle N^{B}\rangle/2=\langle J_{T}^{B}\rangle$.
For a spin-$j$ particle, this is just $\langle J_{T}^{B}\rangle=j.$
With this substitution we obtain inequality \eqref{eq:EPR-Bohm_sum_criterion}.

Using again the linear inferences $B_{\mathrm{est}}(A_{k})=-g_{k}A_{k}+\langle B_{k}+g_{k}A_{k}\rangle$
as discussed above Eq. \eqref{eq:mult_collective_var}, we can derive
directly from \eqref{eq:sum_inf_var_criterion_J}, \eqref{eq:sum_inf_var_criterion_x}
and \eqref{eq:sum_inf_var_criterion_arb} the respective criteria
\begin{equation}
\Delta^{2}(g_{x}J_{x}^{A}+J_{x}^{B})+\Delta^{2}(g_{y}J_{y}^{A}+J_{y}^{B})+\Delta^{2}(g_{z}J_{z}^{A}+J_{z}^{B})\geq\frac{\langle N^{B}\rangle}{2},\label{eq:collective_spin_criterion}\end{equation}
\begin{equation}
\Delta^{2}(g_{x}x^{A}+x^{B})+\Delta^{2}(g_{p}p^{A}+p^{B})\geq2,\label{eq:collective_CV_criterion}\end{equation}
and\begin{equation}
\Delta^{2}(g_{1}A_{1}+B_{1})+\Delta^{2}(g_{2}A_{2}+B_{2})\geq|\langle B_{3}\rangle|_{inf}.\label{eq:collective_arb_criterion}\end{equation}
Again we should keep in mind that the corresponding operators for
Alice, and the values of $g_{k}$, are arbitrary, and therefore should
be chosen so as to optimize the violation of the criteria. Inequality
\eqref{eq:collective_CV_criterion}, which was introduced in \citep{Reid2008tb},
is the analogue for EPR-steering of the entanglement criteria of Duan
\emph{et al.} \citep{Duan2000} and Simon \citep{Simon2000}. Note
that the bound is half that of those authors (making it harder to
violate), a consequence of the fact that EPR-steering is a stronger
form of nonlocality than entanglement. Inequality \eqref{eq:collective_spin_criterion}
is the analogue of the separability criteria of \citet{Hofmann2003a}.

The inference variance criteria have an immediate interpretation as
a demonstration of the situation described by EPR, as they are based
on an apparent violation of the uncertainty principle by inference
of the variances of the distant subsystem. However, in general any
constraint that can be derived from the LHS model is an EPR-steering
criterion, and by the arguments of Sections \ref{sec:History-and-concepts}
and \ref{sec:Locality-models}, a demonstration of the EPR paradox.
We present below examples of such more general criteria which can
be derived as special cases of the additive convex criterion \eqref{eq:add_convex_criteria}.

\subsubsection{Examples: linear criteria}

We first illustrate this approach by deriving a simple criteria for
the case of two qubits. We start with a quantum constraint on expectation
values of spin-1/2 observables:\begin{equation}
\langle J_{x}\rangle_{\rho}+\langle J_{y}\rangle_{\rho}\leq\frac{\sqrt{2}}{2}.\label{eq:constraint_Jx_Jy}\end{equation}
This must be satisfied by any quantum state of a qubit: $\frac{1}{\sqrt{2}}(\hat{J}_{x}+\hat{J}_{y})\equiv\hat{J}_{\theta}$
is simply the observable corresponding to the spin projection on a
direction at $\theta=45^{o}$ between $\mathbf{x}$ and $\mathbf{y}$,
and so for any quantum state $\rho$, $\langle\hat{J}_{\theta}\rangle_{\rho}\leq\frac{1}{2}$.

Now it must then also be the case that, for a pair of observables
$\hat{J}_{x}^{B},\,\hat{J}_{y}^{B}$ for Bob and $\hat{J}_{x}^{A},\,\hat{J}_{y}^{A}$
for Alice, and where $\alpha_{i}\in\{-\frac{1}{2},\frac{1}{2}\}$
represent possible values for the outcomes of observable $\hat{J}_{i}^{A},$\begin{equation}
\alpha_{x}\langle J_{x}^{B}\rangle_{\rho}+\alpha_{y}\langle J_{y}^{B}\rangle_{\rho}\leq\frac{\sqrt{2}}{4},\label{eq:bound_spin_1/2}\end{equation}
for all values of $\alpha_{x},\,\alpha_{y}.$ This is easy to see
by noting that the different values of $(\alpha_{x},\,\alpha_{y})$
lead to one of $\mp\frac{1}{2}\langle J_{x}^{B}\pm J_{y}^{B}\rangle,$
and for each of these the argument of the previous paragraph leads
to \eqref{eq:bound_spin_1/2}. This is of the form \eqref{eq:add_conv_constraint},
and therefore, by substituting on \eqref{eq:add_convex_criteria}
and noting that $\sum_{A}P(A)\, J_{i}^{A}\langle J_{i}^{B}\rangle_{A}=\langle J_{i}^{A}J_{i}^{B}\rangle,$
it leads to the EPR-steering criterion\begin{equation}
\langle J_{x}^{A}J_{x}^{B}\rangle+\langle J_{y}^{A}J_{y}^{B}\rangle\leq\frac{\sqrt{2}}{4}.\label{eq:two_qubit_criterion_0}\end{equation}
Following a similar procedure, and using the quantum constraint $\alpha_{x}\langle J_{x}^{B}\rangle_{\rho}+\alpha_{y}\langle J_{y}^{B}\rangle_{\rho}\geq-\frac{\sqrt{2}}{4},$
which is valid for the same reason as \eqref{eq:bound_spin_1/2},
we can derive the inequality $\langle J_{x}^{A}J_{x}^{B}\rangle+\langle J_{y}^{A}J_{y}^{B}\rangle\geq-\frac{\sqrt{2}}{4}.$
These two inequalities can be summarised in the EPR-steering criterion
\begin{equation}
\left|\langle J_{x}^{A}J_{x}^{B}\rangle+\langle J_{y}^{A}J_{y}^{B}\rangle\right|\leq\frac{\sqrt{2}}{4}.\label{eq:two_qubit_criterion_1}\end{equation}

A similar, more powerful inequality can be derived from the analogous
constraint on three observables \begin{equation}
-\frac{\sqrt{3}}{2}\leq\alpha_{x}\langle J_{x}\rangle_{\rho}+\alpha_{y}\langle J_{y}\rangle_{\rho}+\alpha_{z}\langle J_{z}\rangle_{\rho}\leq\frac{\sqrt{3}}{2},\label{eq:constraint_Jx_Jy_Jz}\end{equation}
which follows, as \eqref{eq:bound_spin_1/2}, from the fact that $\hat{J}_{\phi}\equiv\frac{1}{\sqrt{3}}(\hat{J}_{x}+\hat{J}_{y}+\hat{J}_{z})$
is another observable corresponding to a spin projection. From \eqref{eq:constraint_Jx_Jy_Jz}
we can derive, following similar steps as above, the EPR-steering
criterion\begin{equation}
\left|\langle J_{x}^{A}J_{x}^{B}\rangle+\langle J_{y}^{A}J_{y}^{B}\rangle+\langle J_{z}^{A}J_{z}^{B}\rangle\right|\leq\frac{\sqrt{3}}{4}.\label{eq:two_qubit_criterion_2}\end{equation}

We can now generalize this to an arbitrary total spin. For a spin-$j$
particle, the quantum constraint $\left|\alpha_{x}\langle J_{x}\rangle_{\rho}+\alpha_{y}\langle J_{y}\rangle_{\rho}+\alpha_{z}\langle J_{z}\rangle_{\rho}\right|\leq\sqrt{3}j^{2}$
holds. To see this, note that $\hat{J}_{\phi}\equiv(\alpha_{x}\hat{J}_{x}+\alpha_{y}\hat{J}_{y}+\alpha_{z}\hat{J}_{z})/\sqrt{\alpha_{x}^{2}+\alpha_{y}^{2}+\alpha_{z}^{2}}$
is again a spin projection operator, and that $\sqrt{\alpha_{x}^{2}+\alpha_{y}^{2}+\alpha_{z}^{2}}\leq\sqrt{3}j.$
Following the same steps as for the derivation of \eqref{eq:two_qubit_criterion_1}
this leads to the EPR-steering inequality\begin{equation}
\left|\langle J_{x}^{A}J_{x}^{B}\rangle+\langle J_{y}^{A}J_{y}^{B}\rangle+\langle J_{z}^{A}J_{z}^{B}\rangle\right|\leq\sqrt{3}j^{2}.\label{eq:two_qudit_criterion}\end{equation}

\subsubsection{Generalisation for positive operator valued measures (POVMs)}

In all of the above we have assumed that the measurements on Bob's
system can be described by observables, with projection operators
associated to eigenvalues. There is no loss of generality in this
assumption if we allow Bob's system to be supplemented by an ancilla
system, uncorrelated with any other system \citep{Helstrom}. However
it is often convenient to consider generalized measurements, described
by a POVM, that is, a set of positive operators $F_{\mu}$ associated
to measurement outcomes $\mu$, which sum to unity. In terms of finding
appropriate EPR-steering criteria, the additive convex criteria are
the ones most naturally generalizable to this case. We replace the
$f_{j}(\langle B_{j}\rangle,\alpha_{j})$ in Eq. \eqref{eq:add_conv_constraint}
by\[
f_{j}(\{\langle F_{\mu}^{j}\rangle_{\rho}:\mu\},\alpha_{j}),\]
 where for all $j$ and $\mu$, $F_{\mu}^{j}\geq0$, and for all $j$,
$\sum_{\mu}F_{\mu}^{j}=1$.

The convexity requirement in $\langle B_{j}\rangle_{\rho}$ would
be replaced by a more general convexity requirement, that for all
$j$ and $\alpha_{j}$, all $\rho$ and $\rho'$, and $0\leq p\leq1$,

\begin{multline}
f_{j}(\{\langle F_{\mu}^{j}\rangle_{\rho''}:\mu\},\alpha_{j})\\
\leq pf_{j}(\{\langle F_{\mu}^{j}\rangle_{\rho}:\mu\},\alpha_{j})+(1-p)f_{j}(\{\langle F_{\mu}^{j}\rangle_{\rho'}:\mu\},\alpha_{j}),\label{eq:POVM_convexity}\end{multline}
where $\rho''=p\rho+(1-p)\rho'$. The derivation of Eq. \eqref{eq:add_convex_criteria}
then follows exactly as before.

\section{Applications to classes of quantum states \label{sec:Applications-to-classes}}

We now apply the criteria derived in the previous section to some
classes of quantum states of experimental interest. Violations of
those inequalities amount to demonstrations of the effect termed {}``steering''
by Schrödinger in his response to EPR, reviewed in Sec.~\ref{sub:Schr=0000F6dinger's-response}.
In the continuous variables case, this provides a more modern and
unifying approach to the demonstration of the correlations considered
by EPR in their original example, discussed in Sec.~\ref{sec:The-Einstein-Podolsky-Rosen-argument}.
In the discrete variables case this represents a modern approach to
the demonstration of EPR-Bohm correlations discussed in Sec.~\ref{sub:Bohm's-version}.
We consider each case in turn.

\subsection{Continuous variables}

\begin{figure}
\begin{centering}
\includegraphics[width=8cm]{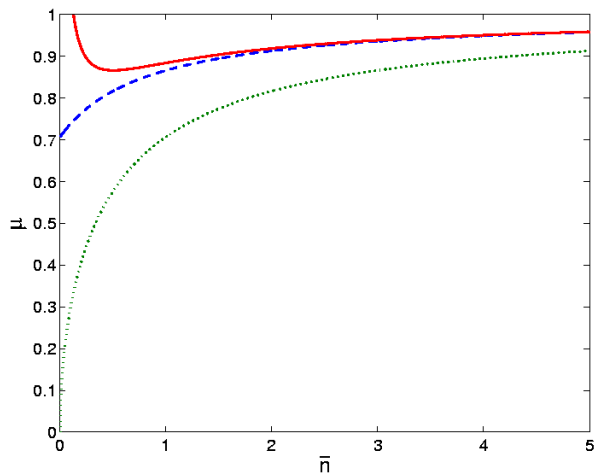}
\par\end{centering}

\caption{\label{fig:CVbounds} (Color on-line.) Boundaries between different
classes of symmetric two-mode Gaussian states. The lower line (green,
dotted) is an entanglement boundary given by Eq.~\eqref{eq:Simon}:
states above the line are entangled. The central (blue, dashed) line
is a steerability (lower) boundary based on Eq.~\eqref{eq:EPRtwomodesteer}
for the EPR paradox: states above this line are steerable. The upper
line (red, full) is a second steerability (lower) boundary based on
a generalisation of the entanglement criterion of Duan \emph{et al.}
\citep{Duan2000} and Simon \citep{Simon2000}: states above this
line are steerable.}

\end{figure}

We consider as a continuous variable example the case of two-mode
Gaussian states prepared by optical parametric amplifiers \citep{Bowen2004}.
Such states include the original EPR state as a special case with
zero entropy and infinite energy. We define $\hat{x}^{A}=\hat{a}+\hat{a}^{\dagger}$
and $\hat{p}^{A}=-i(\hat{a}-\hat{a}^{\dagger})$ as the position and
momentum observables to be measured by Alice, where $\hat{a}$ and
$\hat{a}^{\dagger}$ are the annihilation and creation operators for
a bosonic field mode at Alice's subsystem. We define $\hat{x}^{B},\,\hat{p}^{B}$
analogously for Bob's subsystem in terms of the annihilation and creation
operators $\hat{b}$ and $\hat{b}^{\dagger}$ for his field mode.
When the entanglement is symmetric between the two modes the covariance
matrix describing such states has a particularly simple form. The
continuous variable entanglement properties of such a state have recently
been characterized experimentally \citep{Bowen2004}.. In this case
the covariance matrix of the state $W$ has just two parameters, $\mu$
and $\bar{n}$:\begin{equation}
{\rm CM}[W_{\bar{n}}^{\mu}]=V_{2}^{\alpha\beta}=\left(\begin{array}{cccc}
\gamma & 0 & \delta & 0\\
0 & \gamma & 0 & -\delta\\
\delta & 0 & \gamma & 0\\
0 & -\delta & 0 & \gamma\end{array}\right),\label{eq:GaussCovariance}\end{equation}
where $\gamma=1+2\bar{n}$ and $\delta=2\eta\sqrt{\bar{n}(1+\bar{n})}$.
Here $\bar{n}$ is the mean photon number for each party, and $\mu$
is a mixing parameter defined such that the covariance matrix is linear
in $\mu$ and that $0\leq\mu\leq1$, such that $\mu=0$ corresponds
to an uncorrelated state and $\mu=1$ corresponds to a pure state
\citep{Jones2007}. It has been shown by Duan \emph{et al.} \citep{Duan2000}
and Simon \citep{Simon2000} that if a quantum state such as $W_{\bar{n}}^{\mu}$
is separable it must satisfy\begin{equation}
\Delta^{2}(x^{A}-x^{B})+\Delta^{2}(p^{A}+p^{B})\geq4.\label{eq:Simon}\end{equation}
It is straightforward to show that for states defined by Eq.~\eqref{eq:GaussCovariance}
this leads to the condition that\begin{equation}
\mu>\frac{\bar{n}}{\sqrt{\bar{n}(1+\bar{n})}}\end{equation}
indicates entanglement. This condition is plotted in Fig.~\ref{fig:CVbounds},
where states above the line are entangled.

As discussed in Sec.~\ref{sec:Experimental-criteria-for}, the generalization
of Duan \emph{et al.} and Simon's entanglement criterion to EPR-steering
is given by inequality \eqref{eq:collective_CV_criterion}. For states
of the form of Eq.~\eqref{eq:GaussCovariance}, the relevant criterion
becomes, using the optimal scale factors $g_{x}=-1$ and $g_{p}=1$,\begin{equation}
\Delta^{2}(x^{A}-x^{B})+\Delta^{2}(p^{A}+p^{B})\geq2.\label{eq:SimonLike}\end{equation}
For the two-mode symmetric states we find\begin{equation}
\Delta^{2}(x^{A}-x^{B})=\Delta^{2}(p^{A}+p^{B})=2\gamma-2\delta.\end{equation}
Substituting into \eqref{eq:SimonLike} and rearranging we find that\begin{equation}
\mu>\frac{1+4\bar{n}}{4\sqrt{\bar{n}(1+\bar{n})}}\label{eq:SimonSteer}\end{equation}
indicates EPR-steering. This condition is plotted in Fig.~\ref{fig:CVbounds},
where states above the line are steerable. For this particular state
the additive convex criterion \eqref{eq:SimonLike} and the corresponding
multiplicative criterion\begin{equation}
\Delta^{2}(x^{A}-x^{B})\Delta^{2}(p^{A}+p^{B})\geq1,\end{equation}
derived from \eqref{eq:mult_collective_var}, give the same results,
since both variances are identical in this case.

For comparison, recall the EPR-Reid criterion, \eqref{eq:EPR-Reid_rederived},
which tells us that the violation of\begin{equation}
\Delta_{\mathrm{inf}}x^{B}\Delta_{\mathrm{inf}}p^{B}\geq1\label{eq:EPRsteer}\end{equation}
indicates EPR-steering. Evaluating the left hand side of \eqref{eq:EPRsteer}
for two-mode symmetric Gaussian states, using the optimal inference
variances $\Delta_{\mathrm{min}}x^{B}$ as defined in Eq. \eqref{eq:min_inf_var},
we thus obtain\begin{equation}
\mu>\sqrt{\frac{1+2\bar{n}}{2(1+\bar{n})}}\label{eq:EPRtwomodesteer}\end{equation}
as a condition indicating the demonstration of EPR-steering. Also
in this case inequality \eqref{eq:EPRsteer} detects EPR-steering
just as well as the analogous additive criterion \eqref{eq:sum_inf_var_criterion_x},
since both inference variances for $x^{B}$ and $p^{B}$ have the
same value. In Fig.~\ref{fig:CVbounds} we see that \eqref{eq:EPRsteer}
provides a lower bound on steerability than that provided by \eqref{eq:SimonLike}
(although for $\bar{n}\gg1$ the two bounds become arbitrarily close).
This is not surprising when one remembers, as discussed in Sec.~\ref{sub:Multiplicative-variance-criteria},
that the optimal conditional variances \eqref{eq:EPRsteer} are lower
bounds for the linear-estimate inference of the form $\Delta^{2}(g_{x}x^{A}+x^{B})$.
In other words, as pointed out in Sec.~\ref{sec:Experimental-criteria-for},
the EPR criterion is a more sensitive witness to EPR-steering than
inequality \eqref{eq:SimonLike}, derived as the steerability generalisation
of the entanglement criterion of Duan \emph{et al.} and Simon.

\subsection{Discrete variables}

To illustrate the use of EPR-steering criteria in the discrete variable
case we will make use of the Werner states \citep{Werner1989}. For
the case of a two-dimensional subsystems, these are a natural mixed-state
generalization of the singlet state considered by Bohm, and can be
written as follows 

\begin{equation}
\rho_{W}=\mu|\psi_{S}\rangle\langle\psi_{S}|+(1-\mu)\frac{\mathbf{I}}{4},\label{eq:WernerState}\end{equation}
where $|\psi_{S}\rangle=\frac{1}{\sqrt{2}}(|\frac{1}{2}\rangle|-\frac{1}{2}\rangle-|-\frac{1}{2}\rangle|\frac{1}{2}\rangle)$,
$\mathbf{I}$ is the identity over both subsystems, and $\mu$ is
a mixing parameter that can take values $\mu\leq1$, with $\mu=0$
again corresponding to a product state \citep{Wiseman2007}. 

It was shown in Ref. \citep{Wiseman2007} that the Werner state is
steerable in theory with an infinite number of measurements whenever
$\mu>1/2$. In order to demonstrate EPR-steering in a realistic experimental
setup it is sufficient to instead test a suitable EPR-steering criterion.

We will first evaluate the criterion given by inequality \eqref{eq:EPR-Bohm_criterion}.
Calculation shows that for the Werner state \eqref{eq:WernerState},\[
\Delta_{\mathrm{inf}}^{2}J_{z}^{B}=\frac{1}{4}(1-\mu^{2})\]
and\[
|\langle J_{z}^{B}\rangle|_{\mathrm{inf}}=\frac{\mu}{2}.\]
The Werner state is rotationally symmetric, and thus $\Delta_{\mathrm{inf}}J_{x}^{B}=\Delta_{\mathrm{inf}}J_{y}^{B}=\Delta_{\mathrm{inf}}^{2}J_{z}^{B}$.
We therefore find that inequality \eqref{eq:EPR-Bohm_criterion} will
be violated (demonstrating EPR-steering) for $\mu>(\sqrt{5}-1)/2\approx0.62$.
This inequality cannot therefore detect all steerable states.

For inequality \eqref{eq:sum_inf_var_criterion_J} we make the substitution
(as explained below Eq. \eqref{eq:sum_inf_var_criterion_J}) $\langle N^{B}\rangle/2=j=1/2$,
and with the values for $\Delta_{\mathrm{inf}}^{2}J_{z}^{B}$ a simple
calculation reveals violation whenever $\mu>1/\sqrt{3}\approx0.58$,
This inequality, more symmetric between the different measurements,
thus detects more steerable states (within the class of Werner states)
than the less symmetric \eqref{eq:EPR-Bohm_criterion}.

We now proceed to evaluating the linear inequalities \eqref{eq:two_qubit_criterion_1}
and \eqref{eq:two_qubit_criterion_2}. The expectation value of the
products of observables required for those inequalities, given the
Werner state, is\[
\langle J_{i}^{A}J_{i}^{B}\rangle=-\frac{\mu}{4},\]
where again by symmetry those expectation values are the same for
all $i\in\{x,y,z\}$. Substituting in \eqref{eq:two_qubit_criterion_1}
we obtain a violation for $\mu>1/\sqrt{2}\approx0.71$ and in \eqref{eq:two_qubit_criterion_2},
violation for $\mu>1/\sqrt{3}\approx0.58$. The first inequality,
with only two measurements per site, performs worse (detects less
steerable Werner states) than \eqref{eq:EPR-Bohm_criterion}, but
the second, with three measurements, detects a larger range. Note
that the range of states for which violation is predicted using \eqref{eq:sum_inf_var_criterion_J}
is the same as that detected with \eqref{eq:two_qubit_criterion_2}.
The latter, however, offers the advantage of being simpler to measure
and calculate.

\section{Conclusion\label{sec:Discussion-and-conclusions}}

We have developed a general theory of EPR-steering criteria. These
criteria are the experimental consequences of a LHS model for one
party (Bob), just as Bell inequalities are the experimental consequence
of a LHV model and entanglement criteria are consequences of a quantum
separable model. The essential ingredients in the derivation of the
criteria are the convexity of the set of correlations that allow a
LHS model and (generalized) uncertainty relations which define bounds
on how Bob's outcomes can be described by quantum states.

Analysing the different forms of nonlocality, we see that they differ
only in how they treat the states of Alice and/or Bob, but they are
all convex combinations of separable probability distributions. Some
of the criteria derived here were therefore similar to known entanglement
criteria, but with a more restrictive bound due to the fact that Alice's
subsystem is treated as an arbitrary hidden-variable state. However
others, in particular the \emph{linear EPR-steering criteria, }are
entirely new. These criteria open the possibility to new experimental
demonstrations of the EPR-steering phenomenon, with close links to
topics in quantum information including entanglement witnesses and
quantum cryptography.
\begin{acknowledgments}
We would like to acknowledge support from the Griffith University
Postdoctoral Fellowship scheme, Australian Research Council grants
FF0458313, DP0984863, the ARC Centre of Excellence for Quantum Computing
Technology and the ARC Centre of Excellence for Quantum-Atom Optics.
\end{acknowledgments}
\bibliographystyle{unsrt}
\bibliography{/Users/eric/Documents/Work/bib/Eric}

\begin{thebibliography}{10}

\bibitem{Einstein1935}
A.~Einstein, B.~Podolsky, and N.~Rosen.
\newblock Can quantum-mechanical description of physical reality be considered
  complete?
\newblock {\em Physical Review}, 47:777, 1935.

\bibitem{Bell1971}
J.~S. Bell.
\newblock Introduction to the hidden-variable question.
\newblock In {\em Foundations of Quantum Mechanics}, page 171, New York, 1971.
  Academic Press.

\bibitem{Reid1989}
M.~D. Reid.
\newblock Demonstration of the {E}instein-{P}odolsky-{R}osen paradox using
  nondegenerate parametric amplification.
\newblock {\em Physical Review A}, 40(2):913--923, 1989.

\bibitem{Wiseman2006}
H.~M. Wiseman.
\newblock From {E}instein's theorem to {B}ell's theorem: a history of quantum
  non-locality.
\newblock {\em Contemporary Physics}, 47(2):79--88, 2006.

\bibitem{Vedral2006}
Vlatko Vedral.
\newblock {\em Introduction to Quantum Information Science}.
\newblock Oxford University Press, New York, 2006.

\bibitem{SchPCP35}
E.~Schrodinger.
\newblock Discussion of probability relations between separated systems.
\newblock {\em Proc. Cambridge Philos. Soc.}, 31:555, 1935.

\bibitem{Bell1964}
J.~S. Bell.
\newblock On the {E}instein-{P}odolsky-{R}osen paradox.
\newblock {\em Physics}, 1:195, 1964.

\bibitem{Bohm1951}
D.~Bohm.
\newblock {\em Quantum Theory}, chapter~22.
\newblock Prentice Hall, Englewood Cliffs, N.J., 1951.

\bibitem{Cavalcanti2007b}
E.~G. Cavalcanti and M.~D. Reid.
\newblock Uncertainty relations for the realization of macroscopic quantum
  superpositions and {EPR} paradoxes.
\newblock {\em Journal of Modern Optics}, 54:2373, 2007.

\bibitem{Cavalcanti2009b}
E.~G. Cavalcanti, P.~D. Drummond, H.~A. Bachor, and M.~D. Reid.
\newblock Unambiguous signatures of entanglement and {B}ohm's spin {E}{P}{R}
  paradox.
\newblock {\em Optics Express}, 17(21):18693--702, 2009.
\newblock arxiv:0711.3798v1.

\bibitem{Wiseman2007}
H.~M. Wiseman, S.~J. Jones, and A.~C. Doherty.
\newblock Steering, entanglement, nonlocality, and the einstein-podolsky-rosen
  paradox.
\newblock {\em Physical Review Letters}, 98(14):140402, 2007.

\bibitem{Schroedinger1935}
E.~Schr\"odinger.
\newblock Die gegenw\"artige situation in der quantenmechanick.
\newblock {\em Naturwissenschaften}, 23:807, 1935.

\bibitem{Duan2000}
L.~M. Duan, G.~Giedke, J.~I. Cirac, and P.~Zoller.
\newblock Inseparability criterion for continuous variable systems.
\newblock {\em Physical Review Letters}, 84(12):2722--2725, 2000.

\bibitem{Simon2000}
R.~Simon.
\newblock Peres-{H}orodecki separability criterion for continuous variable
  systems.
\newblock {\em Physical Review Letters}, 84:2726, 2000.

\bibitem{Hofmann2003a}
H.~F. Hofmann and S.~Takeuchi.
\newblock Violation of local uncertainty relations as a signature of
  entanglement.
\newblock {\em Physical Review A}, 68(3):032103, 2003.

\bibitem{Guhne2004a}
O.~Guhne.
\newblock Characterizing entanglement via uncertainty relations.
\newblock {\em Physical Review Letters}, 92(11):117903, 2004.

\bibitem{Clauser1969}
J.~F. Clauser, M.~A. Horne, A.~Shimony, and R.~A. Holt.
\newblock Proposed experiment to test local hidden-variable theories.
\newblock {\em Physical Review Letters}, 23(15):880, 1969.

\bibitem{Mermin1980}
N.~D. Mermin.
\newblock Quantum-mechanics vs local realism near the classical limit: a {B}ell
  inequality for spin-s.
\newblock {\em Physical Review D}, 22(2):356--361, 1980.

\bibitem{Fine1982}
A.~Fine.
\newblock Hidden-variables, joint probability, and the {B}ell inequalities.
\newblock {\em Physical Review Letters}, 48(5):291--295, 1982.

\bibitem{Pitowsky1989}
I.~Pitowsky.
\newblock {\em Quantum Probability - Quantum Logic}, volume 321 of {\em Lecture
  Notes in Physics}.
\newblock Springer-Verlag, 1989.

\bibitem{Ardehali1992}
M.~Ardehali.
\newblock {B}ell inequalities with a magnitude of violation that grows
  exponentially with the number of particles.
\newblock {\em Physical Review A}, 46(9):5375--5378, 1992.

\bibitem{Belinskii1993}
A.~V. Belinskii and D.~N. Klyshko.
\newblock Iinterference of light and {B}ell's theorem.
\newblock {\em Physics-Uspekhi}, 36:653, 1993.

\bibitem{Peres1999}
A.~Peres.
\newblock All the {B}ell inequalities.
\newblock {\em Foundations of Physics}, 29(4):589--614, 1999.

\bibitem{Werner2001}
R.~F. Werner and M.~M. Wolf.
\newblock All-multipartite {B}ell-correlation inequalities for two dichotomic
  observables per site.
\newblock {\em Physical Review A}, 64(3):032112, 2001.

\bibitem{Collins2002a}
D.~Collins, N.~Gisin, N.~Linden, S.~Massar, and S.~Popescu.
\newblock {B}ell inequalities for arbitrarily high-dimensional systems.
\newblock {\em Physical Review Letters}, 88(4):040404, 2002.

\bibitem{Zukowski2002a}
M.~Zukowski and C.~Brukner.
\newblock {B}ell's theorem for general n-qubit states.
\newblock {\em Physical Review Letters}, 88(21):210401, 2002.

\bibitem{Cavalcanti2007a}
E.~G. Cavalcanti, C.~J. Foster, M.~D. Reid, and P.~D. Drummond.
\newblock Bell inequalities for continuous-variable correlations.
\newblock {\em Physical Review Letters}, 99:210405, 2007.

\bibitem{Jones2007}
S.~J. Jones, H.~M. Wiseman, and A.~C. Doherty.
\newblock Entanglement, {E}instein-{P}odolsky-{R}osen correlations, {B}ell
  nonlocality, and steering.
\newblock {\em Physical Review A}, 76:052116, 2007.

\bibitem{Brassard2006}
G.~Brassard and A.~A. M\'{e}thot.
\newblock Can quantum-mechanical description of physical reality be considered
  incomplete?
\newblock {\em International Journal of Quantum Information}, 4(1):45--54,
  2006.

\bibitem{SchPCP36}
E.~Schrodinger.
\newblock Probability relations between separated systems.
\newblock {\em Proc. Cambridge Philos. Soc.}, 32:446, 1936.

\bibitem{Gisin1991}
N.~Gisin.
\newblock {B}ell inequality holds for all non-product states.
\newblock {\em Physics Letters A}, 154(5-6):201--202, 1991.

\bibitem{Bohm1952a}
David Bohm.
\newblock A suggested interpretation of the quantum theory in terms of "hidden"
  variables. i.
\newblock {\em Physical Review}, 85:166 -- 179, 1952.

\bibitem{Bohm1952b}
David Bohm.
\newblock A suggested interpretation of the quantum theory in terms of "hidden"
  variables. ii.
\newblock {\em Physical Review}, 85:180 -- 193, 1952.

\bibitem{Bleuler1948}
E.~Bleuler and H.~L. Bradt.
\newblock Correlation between the states of polarization of the two quanta of
  annihilation radiation.
\newblock {\em Physical Review}, 73:1398, 1948.

\bibitem{Wu1950}
C.~S. Wu and I.~Shaknov.
\newblock The angular correlation of scattered annihilation radiation.
\newblock {\em Physical Review}, 77(1):136--136, 1950.

\bibitem{Kocher1967}
Carl~A. Kocher and Eugene~D. Commins.
\newblock Polarization correlation of photons emitted in an atomic cascade.
\newblock {\em Phys. Rev. Lett.}, 18(15):575--577, 1967.

\bibitem{Furry1936}
W.~H. Furry.
\newblock Note on the quantum-mechanical theory of measurement.
\newblock {\em Physical Review}, 49(5):393--399, 1936.

\bibitem{Reid2008tb}
M.~D. Reid, P.~D. Drummond, W.~P. Bowen, E.~G. Cavalcanti, P.~K. Lam, H.~A.
  Bachor, U.~L. Andersen, and G.~Leuchs.
\newblock Colloquium: The {E}instein-{P}odolsky-{R}osen paradox: From concepts
  to applications.
\newblock arXiv:0806.0270, Rev. Mod. Phys., in print, 2009.

\bibitem{Ou1992}
Z.~Y. Ou, S.~F. Pereira, H.~J. Kimble, and K.~C. Peng.
\newblock Realization of the {E}instein-{P}odolsky-{R}osen paradox for
  continuous-variables.
\newblock {\em Physical Review Letters}, 68(25):3663--3666, 1992.

\bibitem{Zhang2000}
Y.~Zhang, H.~Wang, X.~Y. Li, J.~T. Jing, C.~D. Xie, and K.~C. Peng.
\newblock Experimental generation of bright two-mode quadrature squeezed light
  from a narrow-band nondegenerate optical parametric amplifier.
\newblock {\em Physical Review A}, 62(2):023813, 2000.

\bibitem{Silberhorn2001}
C.~Silberhorn, P.~K. Lam, O.~Weiss, F.~Konig, N.~Korolkova, and G.~Leuchs.
\newblock Generation of continuous variable {E}instein-{P}odolsky-{R}osen
  entanglement via the {K}err nonlinearity in an optical fiber.
\newblock {\em Physical Review Letters}, 86(19):4267--4270, 2001.

\bibitem{Schori2002}
C.~Schori, J.~L. Sorensen, and E.~S. Polzik.
\newblock Narrow-band frequency tunable light source of continuous quadrature
  entanglement.
\newblock {\em Physical Review A}, 66:033802, 2002.

\bibitem{Bowen2003}
W.~P. Bowen, R.~Schnabel, P.~K. Lam, and T.~C. Ralph.
\newblock Experimental investigation of criteria for continuous variable
  entanglement.
\newblock {\em Physical Review Letters}, 90(4):043601, 2003.

\bibitem{Howell2004}
J.~C. Howell, R.~S. Bennink, S.~J. Bentley, and R.~W. Boyd.
\newblock Realization of the {E}instein-{P}odolsky-{R}osen paradox using
  momentum- and position-entangled photons from spontaneous parametric down
  conversion.
\newblock {\em Physical Review Letters}, 92(21):210403, 2004.

\bibitem{Bell1987}
J.~S. Bell.
\newblock {\em Speakable and Unspeakable in Quantum Mechanics}.
\newblock Cambridge University Press, Cambridge, 1987.

\bibitem{Cavalcanti2010tb}
E.~G. Cavalcanti and H.~M. Wiseman.
\newblock to be published.

\bibitem{Einstein1947}
A.~Einstein.
\newblock Letter to born, 3 march 1947.
\newblock In M.~Born, editor, {\em The {E}instein-{B}orn Letters}, page 158.
  Macmillan, London, 1971.

\bibitem{Cohen-Tannoudji1977}
C.~Cohen-Tannoudji, B.~Diu, and F.~Lal\"oe.
\newblock {\em Quantum Mechanics}, volume~1.
\newblock Wiley Interscience, 1977.

\bibitem{Helstrom}
C.~W. Helstrom.
\newblock {\em Quantum Detection and Estimation Theory}, volume 123 of {\em
  Mathematics in Science and Engineering}.
\newblock Academic Press, New York, 1976.

\bibitem{Bowen2004}
W.~P. Bowen, R.~Schnabel, P.~K Lam, and T.~C. Ralph.
\newblock Experimental characterization of continuous-variable entanglement.
\newblock {\em Phys. Rev. A}, 69(1):012304, 2004.

\bibitem{Werner1989}
R.~F. Werner.
\newblock Quantum states with {E}instein-{P}odolsky-{R}osen correlations
  admitting a hidden-variable model.
\newblock {\em Physical Review A}, 40(8):4277--4281, 1989.

\end{thebibliography}

\end{document}